\documentclass[twocolumn,aps,prl,superscriptaddress]{revtex4}
\usepackage{amssymb}
\usepackage{amsmath}
\usepackage{color}
\usepackage{graphicx}
\usepackage{marvosym}
\usepackage{ulem}
\usepackage{notoccite}

\usepackage{comment}

\usepackage{bbm}

\usepackage{graphicx,tabularx}
\usepackage{dcolumn}
\usepackage{bm}
\usepackage{xcolor}

\usepackage{marvosym}

\usepackage{titlesec}

\newcommand{\be}{\begin{equation}}
\newcommand{\ee}{\end{equation}}

\newcommand{\la}{\langle}
\newcommand{\ra}{\rangle}

\renewcommand{\Re}{{\rm Re}\,}
\renewcommand{\vec}[1]{{\bf #1}}

\setcitestyle{super}

\renewcommand{\bibnumfmt}[1]{(#1)}

\begin{document}
\title{Atomic configuration controlled photocurrent in van der Waals homostructures}
\author{Ying Xiong}
\affiliation{Division of Physics and Applied Physics, Nanyang Technological University, Singapore 637371}
\author{Li-kun Shi}
\affiliation{Division of Physics and Applied Physics, Nanyang Technological University, Singapore 637371}
\affiliation{Max Planck Institute for the Physics of Complex Systems, 01187 Dresden, Germany}
\author{Justin C.W. Song}
\email{justinsong@ntu.edu.sg}
\affiliation{Division of Physics and Applied Physics, Nanyang Technological University, Singapore 637371}

\begin{abstract}
Conventional photocurrents at a p-n junction depend on macroscopic built-in fields and are typically insensitive to the microscopic details of a crystal's atomic configuration. Here we demonstrate how atomic configuration can control photocurrent in van der Waals (vdW) materials. In particular, we find bulk shift photocurrents (SPC) can display a rich (atomic) configuration dependent phenomenology that range from contrasting SPC currents for different stacking arrangements in a vdW homostructure (e.g., AB vs BA stacking) to a strong light polarization dependence for SPC that align with crystallographic axes. Strikingly, we find that SPC in vdW homostructures
can be directed by modest strain, yielding sizeable photocurrent magnitudes under unpolarized light irradiation and manifesting even in the absence of p-n junctions. These demonstrate that SPC are intimately linked to how the Bloch wavefunctions are embedded in real space, and enables a new macroscopic transport probe (photocurrent) of lattice-scale registration in vdW materials. 
\end{abstract}

\maketitle

\begin{figure} [ht!]
    \centering
    \includegraphics[scale=0.4]
{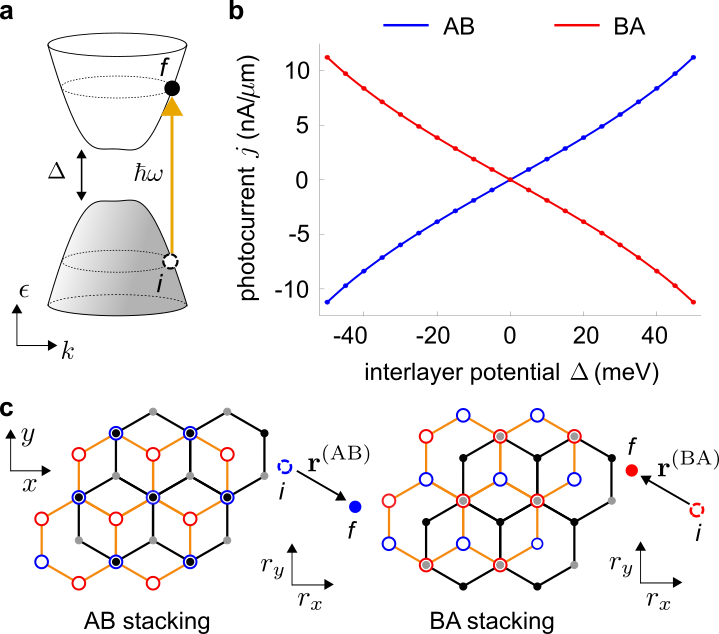}
    \caption{An illustrative example of atomic configuration controlled shift photocurrent (SPC) in vdW homostructures:  
    stacking dependent SPC in bilayer graphene (BLG). (a) Both AB and BA stacking BLG (see panel c) share the same energy spectrum, but have contrasting wavefunctions. This is manifest in the (b) stacking dependence of SPC obtained from Eq.~(\ref{eq:j}) and Eq.~(\ref{eq:H}), flowing in opposite directions for AB vs BA stacking, here $\Delta$ is the interlayer potential. (c) Stacking arrangement of AB and BA stacking BLG; the orange lattice denotes the top layer, the black lattice denotes the bottom layer, open blue and red circles denote A and B sites in the top layer, and filled grey and black circles denote A and B sites in the bottom layer. (inset) the real-space shift that an electron undergoes from $i$ to $f$ states (see panel a) in AB and BA stacking are opposite. Parameters used: $\hbar \omega = 0.1 \; {\rm eV}$, $v = 9.7\times 10^5 \; {\rm m /s}$, $v_3 = 1.2 \times 10^5 \; {\rm m/ s}$, $\gamma_1 = 0.39 \; {\rm eV}$ \cite{BLG_DFT, BLG_parameter1, BLG_parameter2, graphite_parameter}, and $x$-polarised electric field with $E = 0.2 \; {\rm V / \mu m}$. We have fixed $T=10 \; {\rm K}$ and $\mu = 0$.} 
    \label{fig1}
\end{figure}

The atomic scale registration formed when two van der Waals (vdW) layers are stacked on top of each other can have a profound influence on its electronic behaviour~\cite{Geim,EQMM}. Prime examples include strongly correlated phases in moir\'e superlattices \cite{Herrero_TBGMott, Herrero_TBGsuperc,Balents2020}, constructing topologically non-trivial bands from topologically trivial materials through stacking \cite{Song2015, Yao2017}, as well as new types of collective modes (e.g., moir\'e excitons in twisted transition metal dichalcogenides (TMDs)~\cite{Yu2017,Tran2019,Jin2019,Evgeny2019}). Yet in many instances, identifying the stacking arrangement can be challenging since multiple stacking configurations can yield the same electronic energy dispersion, for e.g., AB and BA stacking alignments in vdW bilayers possess the same energy dispersion [Fig.~\ref{fig1}], but contrasting atomic registrations and Bloch wavefunctions.

\begin{figure} [t]
    \centering
    \includegraphics[width=0.95\columnwidth]{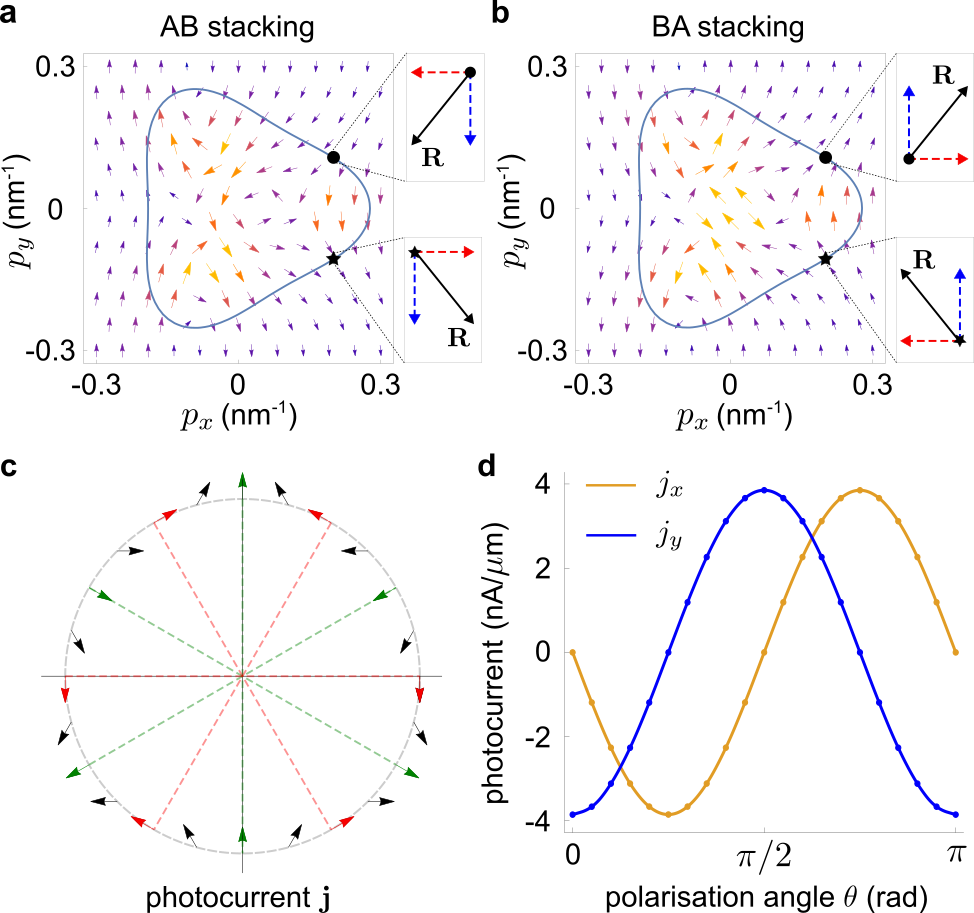}
    \caption{(a, b) Plot of weighted shift vector $\vec R^{\rm (\eta)} (\Delta, \theta =0, \vec p)$ in AB (a) and BA (b) stacked BLG at the $K_+$ valley for $x$-polarised electric field. Here we have taken $\Delta = 20 \; {\rm meV}$ as an illustration; other parameters are the same as Fig.~\ref{fig1}. The solid blue curve denotes the iso-energy contour for the interband transition. (c, d) Polar (c) and line (d) plot of the $x$ (yellow) and $y$ (blue) components of the shift current as a function of electric field polarisation angle $\theta$ with respect to the $x$-axis. The electric field strength amplitude is fixed at $0.2 \; {\rm V /\mu m}$. The arrows indicate the direction of the shift current, which is transverse when light polarisation, $\hat{\vec e}_\theta$, is perpendicular to the mirror axes (dashed red lines) and longitudinal when $\hat{\vec e}_\theta$ is parallel to the mirror axes (dashed green lines). }
    \label{fig2}
\end{figure}

\begin{figure}[t!]
    \centering
    \includegraphics[width=\columnwidth]{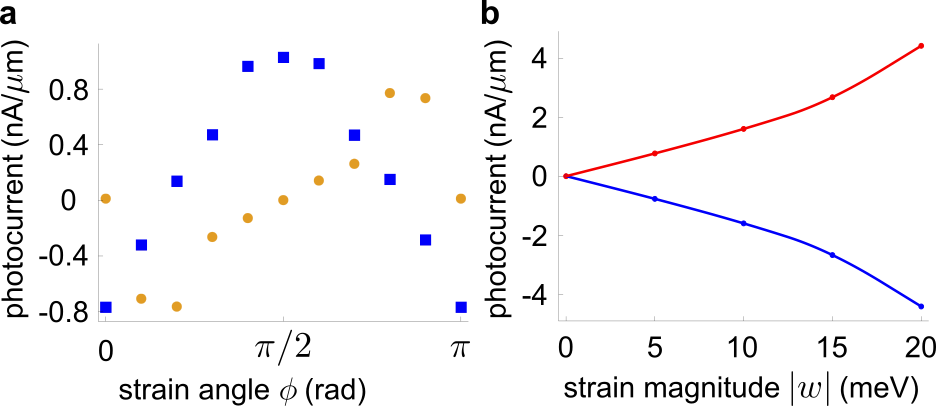}
    \caption{(a) The $x$ (yellow) and $y$ (blue) components of the shift current induced by unpolarised light as a function of direction of the principle axis of a uniaxial strain. The strain magnitude is fixed at $|w| = 5 \; {\rm meV}$. (b) Shift current induced by unpolarised light as a function of strain magnitude fixing $\phi = 0$ (see text) for AB (blue) and BA (red) stacked BLG. All other parameters are the same as Fig.~\ref{fig2}. }
    \label{fig3}
\end{figure}

Here we show that photocurrents excited in vdW homostructures 
can be highly sensitive to its stacking and local atomic configuration. To demonstrate this, we focus on shift photocurrents (SPC) that can be induced in the {\it absence} of p-n junctions~\cite{Kraut1979, Kraut1981, Sipe2000, Rappe2012, BaTiO3_expt, Perovskite_expt2013} in inversion symmetry broken vdW materials~\cite{Pereira}. In particular, we find that SPC in vdW homostructures  
can be directly controlled by the local atomic configuration: for example, we find that in bilayer graphene, SPC flows in opposite directions for AB vs BA stacking (Fig.~\ref{fig1}), exhibits a strong light-polarization dependence flowing transverse (longitudinal) when polarization is aligned perpendicular (parallel) to a mirror axis (Fig.~\ref{fig2}), and can be directed by a local strain profile (Fig.~\ref{fig3}) yielding sizeable currents even for unpolarized light and modest strain values. While we concentrate on bilayer graphene as an illustrative example, as we argue below, atomic configuration sensitivity extends to a host of other vdW materials. 

The sensitivity SPC displays to local atomic configuration contrasts starkly with that of conventional p-n junction photocurrents (e.g., photovoltaic, photothermoelectric or bolometric effects) that arise in the presence of a built-in field or a bias field~\cite{graphene_pn, Lemme, Gabor2011, bolometric, photoconductivity, MoS2_phototherm, Ferrari2014}. The direction (i.e. sign) of these conventional photocurrents only depends on macroscopic and slowly varying variables, such as a built-in field or Seebeck coefficient; as such these conventional photocurrents display photocurrent sign that is insensitive to registration or band gap sign~\cite{graphene_pn, Lemme, Gabor2011, bolometric, photoconductivity, MoS2_phototherm, Ferrari2014}. In contrast, SPC in gapped bilayer graphene displays a sign that changes with registration and gap (Fig.~\ref{fig1}b) and can be tuned by polarization (Fig.~\ref{fig2}), providing a simple and readily accessible experimental signature. Indeed, as we explain below, SPC's sensitivity to atomic configuration proceeds directly from the photoexcitation process: when an electron is photoexcited (Fig.~\ref{fig1}a,c), it undergoes a real-space displacement, $\vec r_{i\to f}$, that depends on how the Bloch wavefunctions are embedded in real space as encoded in its atomic configuration. This renders SPC a sensitive diagnostic of the atomic registration of vdW homostructures. 

We expect atomic configuration sensitive SPC can be readily found in currently available vdW homobilayers (e.g., gapped bilayer graphene) and other homostructures. For instance, stacking dependence can be probed across stacking faults~\cite{Alden2013}; polarization dependence of SPC can be used to identify crystallographic axes. We note that stacking dependent SPC is particularly pronounced in gapped bilayer graphene due to its strong interlayer hybridization~\cite{FanZhang2013, Koshino_BLGreview} that distinguish the wavefunctions from different stacking configurations (e.g., AB/BA). This stands in contrast to vdW heterobilayers (e.g., MoSe2/WSe2) where interlayer hybridization can be weak~\cite{MoS2-WS2-DFT, TMD_heterobilayer, MoS2-WS2, MoS2-WSe2-STM, MoS2-WS2-STM, MoSe2-WSe2-ARPES}, producing a bandstructure and electronic  wavefunctions that are dominated by other factors, such as their type II band alignment~\cite{MoS2-WS2-DFT, TMD_heterobilayer, MoS2-WS2, MoS2-WSe2-STM, MoS2-WS2-STM, MoSe2-WSe2-ARPES}. As a result, in what follows, we will focus on vdW homostructures, and gapped bilayer graphene in particular where interlayer coupling is strong, to exemplify the atomic configuration SPC.

{\it Shift vector and shift photocurrent ---}
We begin by examining the form of SPC that arises from the real space displacement $\vec r_{v \to c}$ that an electron undergoes as it is photo-excited from the valence ($v$) to conduction ($c$) band \cite{Kraut1979,Kraut1981,Sipe2000, Nagaosa}: $\vec j = e \sum_{v\to c} W_{v \to c} \vec r_{v \to c}$, where $W_{v\to c}$ is the rate of photo-excitation from the $v$ to $c$ bands. For vertical transitions, this displacement is described by
a shift vector that depends on the electronic wavefunction in $c,v$ bands \cite{Sipe2000, Nagaosa}: 
\be \label{eq:shiftvector}
\vec r (\theta, \vec k) = \vec A_c (\vec k) - \vec A_v (\vec k) - \nabla_{\vec k} \arg [\nu_\theta (\vec k)],
\ee 
where $\vec k$ is the wavevector of the electron, $\vec A_{c(v)} (\vec k) =  i\la u_{c(v)} (\vec k) |\nabla_{\vec k} | u_{c (v)} (\vec k) \ra$ is the Berry connection of the conduction (valence) band, $\nu_\theta (\vec k)  = \la u_c (\vec k) | \hat{\vec e}_\theta \cdot \hat{\nu} | u_v (\vec k)\ra$ is the velocity matrix element with $\hat{\nu} = \partial_{\vec k} H(\vec k) /\hbar$ the velocity operator, $u_{c(v)} (\vec k)$ is the Bloch wavefunction and $\hat{\vec e}_\theta$ is the incident light electric field polarisation oriented $\theta$ away from $x$-axis. $H(\vec k)$ is the Bloch hamiltonian; here we use $\vec k$ measured away from the $\Gamma$ point. 

Using Eq.~(\ref{eq:shiftvector}), SPC can be written as \cite{Sipe2000, Nagaosa} 
\be \label{eq:j}
\vec j= C \int d^2 \vec k \rho (\vec k) \vec R(\theta, \vec k), \quad \vec R(\theta, \vec k) = |\nu_\theta|^2 \vec r(\theta, \vec k),
\ee
where $C = (e/8 \pi) (eE/\hbar \omega)^2 $, $E$ is the amplitude of the electric field strength of the incident light, $\omega$ is the light frequency. The factor $\rho (\vec k) = (f (\epsilon_{v \vec k}) - f (\epsilon_{c \vec k})) \delta (\omega_{cv} - \omega) $ defines the iso-energy contour that satisfies energy conservation, and $f (\epsilon_{c(v) \vec k})$ is the Fermi-Dirac distribution of electrons.

SPC in Eq.~(\ref{eq:j}) is particularly sensitive to the symmetries of the electronic system manifest in the shift vector. For instance, in the presence of inversion symmetry, the shift vector $\vec r (\theta, \vec k) = - \vec r (\theta, - \vec k)$ is odd. As a result, when there is inversion symmetry, Eq.~(\ref{eq:j}) vanishes \cite{Nagaosa}; breaking inversion symmetry is required for a finite shift current.

{\it Symmetry, configuration, and stacking dependence ---} 
As we now explain, the shift vector is highly sensitive to the atomic configuration. In so doing, we concentrate on stacked bilayers, and introduce a stacking index $\eta = \{ {\rm AB, BA} \}$ to describe the stacking configuration (e.g.  found in BLG) as well as an interlayer potential difference $\Delta$ between top and bottom layers. For concreteness and clarity of presentation, in the main text we will concentrate on BLG where the electronic excitations can be treated as effectively spinless particles. We emphasize that our conclusions and analysis applies more broadly to other vdW materials and homostructures such as monolayer TMDs, bilayer 2H TMDs, as well as gapped graphene on hexagonal Boron Nitride (G/hBN), see discussion in Supplementary Information ({\bf SI})~\cite{SI}.

We proceed by analyzing the crystalline symmetries of Bernal stacked gapped BLG~\cite{BLG_symmetry1, BLG_symmetry2, BLG_symmetry3, Koshino_BLGreview}: they possess a three-fold in-plane rotation $C^z_3$ symmetry and mirror symmetry (MS) about axes that connect the non-dimer sites [Fig.~\ref{fig1}c]. For example, MS about the $y$-axis demands $\mathcal M_y H^{(\eta)} (\Delta, \vec k) \mathcal M_y^{-1} = H^{(\eta)} (\Delta,\mathcal M_y \vec k)$; here $\mathcal{M}_y$ is the mirror operation about the $y$-axis [i.e. $(x,y) \to (-x, y)$]. Applying mirror symmetry as well as time-reversal symmetry~\footnote{We note that for a spinless system that possesses time-reversal symmetry, the shift vector obeys $\vec r^{(\eta)} (\Delta,\theta, \vec k) =  \vec r^{(\eta)} (\Delta,\theta, - \vec k)$. See also full discussion in {\bf SI}.}, we find that $\vec r^{(\eta)}(\Delta, \theta, \vec k)$ transforms as 
\begin{align}\label{eq:Mx}
& r_x^{(\eta)} (\Delta,\theta, k_x,k_y) = - r_x^{(\eta)} (\Delta,- \theta, k_x, -k_y), \nonumber \\
& r_y^{(\eta)} (\Delta,\theta, k_x,k_y) = r_y^{(\eta)} (\Delta, - \theta, k_x,-k_y). 
\end{align}
In obtaining Eq.~(\ref{eq:Mx}), we noted that $\vec r^{(\eta)} (\Delta,\theta, \vec k) = \vec r^{(\eta)} (\Delta,\theta +\pi, \vec k)$ since light polarisations along $\hat{\vec e}_\theta$ and $- \hat{\vec e}_\theta$ are equivalent. While we have focussed on MS about the $y$-axis in Eq.~(\ref{eq:Mx}), there are two other mirror axes related to $\mathcal{M}_y$ via $C^z_3$ rotation from the $y$-axis. In a similar fashion, mirror reflection about these directions produce the same shift vector symmetry relations (for parallel and perpendicular components) as Eq.~(\ref{eq:Mx}), see {\bf SI}~\cite{SI}. 

Interestingly, when the light polarization $\hat{\vec e}_\theta$ is directed parallel (perpendicular) to a mirror axis, e.g., $\theta = \pi/2$ ($\theta = 0)$, the shift vector [see Eq.~(\ref{eq:Mx})] acts as pseudovector; $r_x$ flips sign whereas $r_y$ remains constant as $k_y \to -k_y$. As we will see below, this severely constrains the direction of SPC. For example, when $\hat{\vec e}_\theta$ is parallel to a mirror axis (e.g., $\theta = \pi/2$), SPC is purely longitudinal, flowing parallel to the polarization of light. Similarly, when $\hat{\vec e}_\theta$ is perpendicular to a mirror plane (e.g., $\theta =0$), SPC is purely {\it transverse} (perpendicular to $\hat{\vec e}_\theta$). This latter SPC flow is particularly striking since SPC flows transverse to the incident ac (light) electric field, underscoring the geometric origin of the shift photocurrent.

We now move to the symmetry constraints relating the stacking configurations. AB and BA stackings are related by flipping the sample about the y-axis (Fig.~\ref{fig1}c). For example, the real space Hamiltonians of AB and BA stacked bilayer graphene are related by a $\pi$ rotation about the $y$-axis: $C^y_2 \mathcal H^{\rm (AB)} (\Delta, \vec r) (C^y_2)^{-1}  = \mathcal H^{\rm (BA)} (-\Delta, \vec r) $ where $C^y_2: (x, y, z) \to (-x,  y, -z)$ is the rotation operation. As a result, the Bloch Hamiltonian obeys the symmetry constraint $C^y_2 H^{\rm (AB)} (\Delta, \vec k) (C^y_2)^{-1} = H^{\rm (BA)} (-\Delta, C^y_2 \vec k)$ (see {\bf SI}~\cite{SI} for detailed explanation). Using these and Eq.~(\ref{eq:shiftvector}) we find that AB and BA stackings obey the relation: 
\begin{align}\label{eq:shiftvectorc2x}
& r_{x}^{\rm (AB)} (\Delta,\theta, \vec k) = - r_{ x}^{\rm (BA)} (-\Delta,- \theta, C^y_2 \vec k), \nonumber \\
& r_{ y}^{\rm (AB)} (\Delta, \theta, \vec k) =  r_{y}^{\rm (BA)} ( -\Delta, - \theta, C^y_2 \vec k).
\end{align}
In obtaining Eq.~(\ref{eq:shiftvectorc2x}) we have used that the $x$ component of the velocity matrix element switches sign under $C^y_2$ [i.e. $\hat \nu_x \to - \hat \nu_x$]; this is equivalent to mapping $\theta \to \pi-\theta$.

Noting that under an inversion operation, the shift vector transforms as $\vec r^{(\eta)} (\Delta, \theta, \vec k) = -\vec r^{(\eta)} (-\Delta, \theta, -\vec k)$~\cite{SI}, and applying Eq.~(\ref{eq:Mx}) onto Eq.~(\ref{eq:shiftvectorc2x}) we obtain AB/BA stacking dependent shift vectors that have opposite signs:
\be \label{eq:ABshiftvector}
\vec r^{\rm (AB)} (\Delta, \theta, \vec k)  = -\vec r^{\rm (BA)} (\Delta, \theta, \vec k).
\ee
In obtaining Eq.~(\ref{eq:ABshiftvector}), we noted $C^y_2: (k_x, k_y) \to (-k_x, k_y)$, and applied the time-reversal operation. We remark that this stacking dependence can be understood physically from the inversion-symmetry breaking in the unit cell. For a fixed interlayer potential difference, a change in the stacking order from AB to BA switches the directionality of the in-plane dipole between the non-dimer site, as illustrated in Fig. \ref{fig1}c. Similarly, when interlayer potential is flipped (keeping stacking order the same), the in-plane dipole similarly changes sign. Since the shift vector is sensitive to the electric dipole configuration, this leads to a stacking dependent shift vector in Eq.~(\ref{eq:ABshiftvector}). As we will see below, this directly produces the sign flip of SPC shown in Fig.~\ref{fig1}b.

{\it Illustration: gapped bilayer graphene ---} We now turn to exemplify the stacking (and configuration) dependent SPC in a minimal model of an AB/BA stacked material: BLG. Bernal stacked BLG can be described by a four-band minimal model in the basis $\{\psi_{A_b},\; \psi_{B_b},\; \psi_{A_t}\;, \psi_{B_t} \}$, where $A_{t,b},B_{t,b}$ stand for A and B sites on the top and bottom layers respectively. The Hamiltonian for AB/BA stacked BLG \cite{Koshino_BLGreview, Falko_BLG2006} can be described via $H^{(\eta)} =H_0^{(\eta)} + H_{ w}^{(\eta)}$, where $H_0^{(\eta)}$ and $H_{w}^{(\eta)}$ read
\begin{align}\label{eq:H}
&H_0^{(\eta)} = \hbar v (\xi p_x \sigma_x + p_y \sigma_y) \tau_0 + \frac{\Delta}{2} \sigma_0 \tau_z + \frac{\gamma_1}{2}  (\sigma_x \tau_x + \eta \sigma_y \tau_y), \nonumber \\
&H_{w}^{(\eta)} = \frac{\hbar v_3}{2} \left[ \xi p_x (\sigma_x \tau_x - \eta \sigma_y \tau_y) -  p_y (\sigma_y \tau_x + \eta \sigma_x \tau_y) \right]
\end{align} 
where $\vec p = \vec k - \vec K_{\xi}$ is the wavevector measured from $\vec K_{\xi}$ with $\xi=\pm$ denoting the two valleys. The Pauli matrices $\boldsymbol \sigma$ and $\boldsymbol \tau$ label the sublattice and layer degrees of freedom respectively, and $\eta = \pm$ for AB and BA stacking configurations. Here $v$ is the Fermi velocity in each layer, and $\gamma_1$ is the direct interlayer hopping. The term $H_w$ provides a trigonal warping effect of the energy dispersion with $v_3$ describing the interlayer hopping between the nondimer site \cite{Koshino_BLGreview, Falko_BLG2006}. An interlayer potential difference $\Delta$ opens a gap in the system and breaks inversion symmetry.

Even though trigonal warping $v_3 $ is typically an order of magnitude smaller than $v$ \cite{BLG_DFT, BLG_parameter1, BLG_parameter2, graphite_parameter}, it is nevertheless responsible for enforcing the three-fold symmetry of the bilayers. Importantly, trigonal warping $H_w^{(\eta)}$ is necessary to achieve a non-zero shift current. This can be seen by noting that in the absence of trigonal warping, the low energy dispersion in each valley exhibits an emergent $U(1)$ continuous rotational symmetry. For any linear polarisation $\hat{\vec e}_\theta$, reflection symmetries about axes both parallel and perpendicular to $\hat{\vec e}_\theta$ yield a vanishing shift current.

Numerically computing the shift vector in Eq.~(\ref{eq:shiftvector}) using the eigensolutions of Eq.~(\ref{eq:H}) we obtain the stacking dependent (weighted shift vector) $\vec R^{(\eta)} (\Delta, \theta, \vec p) = \vec r^{(\eta)} (\Delta, \theta, \vec p) |\nu_\theta^{(\eta)} (\vec p)|^2$ in Fig.~\ref{fig2}a,b. We note that $|\nu_\theta^{(\eta)} (\vec p)|^2$ and $\rho (\vec p)$ in Eq.~(\ref{eq:j}) depend only on energy dispersion and are independent of stacking or the gap sign. Here we have chosen an $x$-polarized electric field, $\theta =0$ as an illustration, see {\bf SI} for other polarizations. As expected from Eq.~(\ref{eq:ABshiftvector}), $\vec R^{(\eta)}$ in Fig.~\ref{fig2}a,b flips sign when stacking arrangement changes from AB to BA even when the same interlayer potential is applied. As a result, SPC also switches sign when the shift vector flips sign. Numerically integrating Eq.~(\ref{eq:j}) with the weighted shift vector in Fig.~\ref{fig2}a,b over both valleys (and spins) for a chemical potential in the middle of the gap and $T = 10 \; {\rm K}$, we obtain a stacking dependent SPC shown in Fig.~\ref{fig1}b, that flips sign when either stacking arrangement is changed (AB $\to$ BA) or interlayer potential is switched ($\Delta \to - \Delta$). 

We note that the magnitude of SPC increases as $|\Delta|$ increases yielding sizeable photocurrents of order several  ${\rm nA / \mu m}$ in Fig.~\ref{fig1}b for modest light irradiation (see parameters in caption); it vanishes when no interband transitions occur. While we find maximal SPC occur when chemical potential is in the gap (shown in Fig. \ref{fig1}b), sizeable SPC can still manifest when BLG is doped, but diminish when $2 E_F \sim \hbar \omega$ due to Pauli blocking.

As discussed above in Eq.~(\ref{eq:Mx}), the shift vector can exhibit a pseudovector nature. This is displayed in Fig.~\ref{fig2}a,b for $\theta=0$, wherein recalling that $|\nu_\theta^{(\eta)} (\vec p)|^2$ is even, we have $R_x \to - R_x$ as $p_y \to - p_y$ whereras $R_y$ remains unchanged, see e.g., weighted shift vector on solid circle vs star positions in Fig.~\ref{fig2}a,b inset. Here we have noted that by mapping $k_y \to - k_y$ (measured from $\Gamma$),  $p_y \to - p_y$ in the same valley (see {\bf SI}).

This pseudovector nature (shown in Fig.~\ref{fig2}a,b) yields SPC for $\theta=0$ in Eq.~(\ref{eq:j}) that is purely in the $y$-direction -- transverse to $\hat{\mathbf{e}}_{\theta=0}$. This is manifested in Fig.~\ref{fig2}c, where we plot the SPC $\vec j$ as a function of polarization angle $\theta$ where the dashed lines denote magnitude of SPC and the arrows denote its direction. Strikingly, along polarizations perpendicular to a mirror axes ($\theta = 0, \pm 2\pi/3$, red dashed lines), the shift current is purely transverse (red arrows) to the polarization. This underscores the geometrical origin of SPC with a current that is perpendicular to the applied (oscillating) electric field. In contrast, when light polarization is applied {\it parallel} to a mirror axes ($\theta = \pi/2, \pm \pi/6$), SPC is purely longitudinal (green arrows). As a result, SPC alignment with the crystal axes and its polarization dependence (Fig.~\ref{fig2}d) can be used to determine a vdW device's crystallographic orientation.

{\it Strain photocurrent in van der Waals stacks --- } While large stacking and configuration dependent SPC manifest in the presence of polarized light (discussed above), due to the $C_3^z$ symmetry of the Bernal stacked bilayers, photocurrent vanishes for {\it unpolarized} light irradiation; for a detailed discussion, see {\bf SI}~\cite{SI}. However, when an in-plane strain is applied, it breaks the $C_3^z$ symmetry of the Bernal stacked bilayers, and as we now discuss, enables a strain induced SPC even for unpolarized light. We note that 2D materials can accommodate strains from several to $\sim 10$ percent~\cite{Gstrain1, Gstrain2, MoS2Strain, MoS2_bilayer_Strain}; in addition to extrinsically applied strains~\cite{Gstrain1, Gstrain2, MoS2Strain, MoS2_bilayer_Strain}, strain profiles can also be naturally found in many moire material stacks~\cite{Yoo2019}. We note that strain has recently emerged as a tool for controlling bulk photocurrents \cite{flexoPV, Granzow, Tian, Zheng, Wang, Rappe2020}.

To illustrate such a strain-induced photocurrent in BLG, we consider a uniaxial strain applied in BLG that can be described by $\tilde H^{(\eta)} = \tilde H_0^{(\eta)} + \tilde H_w^{(\eta)} + H_s^{(\eta)}$~\cite{Falko_strainBLG2011, Falko_strainBLG2020}, where $\tilde H_0^{(\eta)}$ and $\tilde H_w^{(\eta)}$ can be obtained from Eq.~(\ref{eq:H}) but replacing $\vec p$ to $\tilde{\vec p} = \vec p+ \vec A_0/v$. Here $\vec A_0 = A_0 (\cos2 \phi , -\xi \sin 2 \phi)$ describes a shift of the valleys in $k$-space, where $\phi$ is the angle of the strain principle axis with respect to the $x$-axis. The additional terms $H_s^{(\eta)}$ describes the modification of the skew interlayer coupling by the strain and is given by~\cite{Falko_strainBLG2011, Falko_strainBLG2020} $H_s^{(\eta)} =  \left[ w_x (\sigma_x \tau_x - \eta \sigma_y \tau_y) - w_y (\sigma_y \tau_x + \eta \sigma_x \tau_y) \right]/2$, where $\vec w = |w| (\cos 2 \phi, - \xi \sin 2 \phi)$ accounts for both the (in-plane) uniaxial strain and an interlayer shear associated with it \cite{Falko_strainBLG2011, Falko_strainBLG2020}. In BLG, $2\%$ of strain corresponds to $|w| \approx 19 \; {\rm meV}$ \cite{Falko_strainBLG2011, Falko_strainBLG2020}. We note that the shift of the valleys does not affect the shift current since it involves integration over the entire $k$-space. In the following, we will focus on the effect of $\vec w$.

When BLG is strained, we find that the shift currents at different polarisation angles do not cancel out. Instead, a net shift current (upon integration over all polarisations) manifests -- the strain photocurrent: $\vec J^{\rm tot} = \int  \vec j (\theta) d\theta$. Strain photocurrents are shown in Fig.~\ref{fig3}a,b where we have numerically integrated Eq.~(\ref{eq:j}) together with the eigensolutions of strained BLG above. As expected, strain photocurrents increase with increasing strain (Fig.~\ref{fig3}b), achieving sizeable values of order ${\rm nA / \mu m}$ for modest strain values and incident light irradiation (see parameters in caption). We note that, similar to that discussed for SPC, the strain photocurrent in Fig.~\ref{fig3}b have opposite signs for AB vs BA (blue vs red).

Strain photocurrents are very sensitive to the principle axis ($\phi$) of applied strain and exhibit a $\pi$ periodicity, see Fig.~\ref{fig3}a. Strikingly, when strain principle axis is perpendicular (parallel) to the mirror axes, the strain photocurrent induced is transverse (longitudinal) to the applied strain direction, see e.g. strain photocurrents $\phi =0$ ($\phi = \pi/2$) in Fig.~\ref{fig3}a. As we now explain, this directly proceeds from the symmetry of the bilayer stacks. We write the strain photocurrent in terms of its symmetric and antisymmetric (w.r.t. $\theta$) contributions 
\begin{align}
J_{i}^{\rm tot} =  & C \sum_{\xi =\pm} \int 
 \int d\theta d \vec p \rho(\vec p) [ V_s(\theta, \vec p) ( r_i (\theta, \vec p) + r_i (-\theta, \vec p)) \nonumber \\
&+ V_a (\theta, \vec p) (r_i (\theta, \vec p) - r_i (-\theta, \vec p)) ],
\label{eq:jstrain}
\end{align}
where $V_s (\theta, \vec p) = | \nu_x (\vec p)|^2 \cos^2 \theta  +  | \nu_y (\vec p)|^2 \sin^2 \theta$ is the component of $| \nu_\theta (\vec p)|^2$ that is symmetric in $\theta$, and $V_a (\theta, \vec p) = 2 \Re [ \nu_x (\vec p) \nu_y (\vec p)^*] \sin \theta \cos \theta $ is the component that is antisymmetric in $\theta$. Here $\int d\theta$ denotes an integral of polarization angle between $0\leq \theta \leq\pi/2$ and we have omitted mention of the stacking index $\eta$ and interlayer potential difference, $\Delta$, for brevity. 

Crucially, when strain is applied either parallel or perpendicular to the mirror axes, MS about the mirror axis is preserved. These constrain the integrand of Eq.~(\ref{eq:jstrain}).  For example, when strain is along $\phi=0, \pi/2$, we observe that $V_{s,a} (\theta, \vec p)$ is even (odd) under $p_y \to - p_y$; similarly, the shift vector transforms according to Eq.~(\ref{eq:Mx}). Combining these, we find that the integrand for $J_{x}^{\rm tot}$ is odd under $p_y \to - p_y$, whereas $J_{y}^{\rm tot}$ is even. As a result, when strain is applied along $\phi=0$ ($\phi=\pi/2$), strain photocurrent is purely transverse (longitudinal) as shown in Fig.~\ref{fig3}a. The same reasoning can be applied for strains along other high symmetry axes, see {\bf SI}~\cite{SI}. This demonstrates vividly how strains can be used to direct SPC.

SPC is a quantum geometric property/response that depends on how and where the wavefunctions in the unit cell are embedded in real space. The sensitivity to the real-space embedding is particularly pronounced in gapped bilayer graphene, enabling a wealth of SPC properties that include stacking and light polarization angle dependence,
as well as a sensitivity to strain in the sample. We note that such sensitivity to the atomic registration often requires scanning probe or transmission electron microscopy techniques. SPC, on the other hand, can be readily extracted at global leads in a conventional scanning photocurrent experiment~\cite{Woessner2016,Sunku2020} enabling a new tool for crystallographic and strain profile characterization. Indeed, a particularly urgent venue for such characterization are low twist-angle moire materials (e.g., twisted BLG with twist angle $\lesssim 1^\circ$) wherein atomic reconstruction leads to an alternating pattern of commensurate AB and BA stacking domains with sizes as large as several hundreds of nm~\cite{Yoo2019, Basov2018, McEuen}. In these systems, we expect local SPC induced by light irradiation concentrated in AB vs BA stacking domains will flow in opposite directions. Furthermore, real-space strain profiles may additionally warp the direction of the polarisation dependent SPC. As a result, we  anticipate SPC (aided by scanning near-field optical probes~\cite{Woessner2016,Sunku2020}) can enable macroscopic transport probe of the myriad atomic registrations found in twisted materials. 

From a technological perspective, we note that SPC can attain magnitudes up to tens ${\rm nA /\mu m}$ (for modest irradiation of several ${\rm kW}{\rm / cm^2}$ and $\hbar \omega = 0.1 {\rm  \;eV}$), and are comparable with those typically found in graphene based photodetectors~\cite{Gabor2011}. Indeed, since gap sizes can reach large values, SPC can persist to high temperatures including room temperature. Because SPC in Eq.~(\ref{eq:j}) scales inversely with photon frequency, we anticipate responsivities and photocurrent will be further enhanced as $\omega$ is lowered into the THz regime.

\vspace{2mm}
\begin{acknowledgments}
{\it Acknowledgements -- } We acknowledge insightful conversations with Arpit Arora and Qiong Ma.
This work was supported by the National Research Foundation (NRF), Singapore under its NRF fellowship programme award number NRF-NRFF2016-05, the Ministry of Education, Singapore under its MOE AcRF Tier 3 Award MOE2018-T3-1-002, and a Nanyang Technological University start-up grant (NTU-SUG). 
\end{acknowledgments}

\clearpage

\newpage

\setcounter{equation}{0}
\setcounter{figure}{0}
\renewcommand{\theequation}{S\arabic{equation}}
\renewcommand{\thefigure}{S\arabic{figure}}

\renewcommand{\bibnumfmt}[1]{(#1)}

\setcounter{secnumdepth}{4}

\onecolumngrid

\begin{center}
\textbf{\large Supplementary Information for ``Atomic configuration controlled photocurrent in van der Waals homostructures"} 
\end{center}

\section*{}
\underline{\bf Contents of the Supplementary Information} 
\\\\
I. {\bf Symmetry and Stacking Analysis of Shift Vector in van der Waals materials and homostructures}
\\
\hspace{2mm} A. {\it Shift Vector Configuration Dependence for Bernal Stacked BLG} 
\\
\hspace{2mm} B. {\it Shift Vector Configuration Dependence for Staggered Sublattice Potential, e.g. G/hBN}
\\
\hspace{2mm} C. {\it Shift Vector Configuration Dependence for Monolayer TMDs}
\\
\hspace{2mm} D. {\it Shift Vector Configuration Dependence for 2H Stacked Bilayer TMDs}
\\\\
II. {\bf Strained shift current induced by unpolarized light}
\\\\
III. {\bf Hamiltonian of Bernal Stacked BLG}
\\\\
IV. {\bf Numerical Calculation of Shift Vector in AB stacked BLG for other polarizations}

\section{Symmetry and Stacking Analysis of Shift Vector in van der Waals materials and homostructures}

In this section, we present the shift vector dependence on the symmetry and stacking arrangement in van der Waals (vdW) materials and homostructures.
We show that the shift vector is highly sensitive to the local atomic configuration of the structure, leading to stacking and polarisation dependent SPC. In general, the real space shift of photo-excited electrons can be described by a shift vector $\vec r (\theta, \vec k)$ \cite{Sipe2000, Nagaosa} as displayed in Eq.~(1) of the main text: 
\be 
\vec r (\theta, \vec k) = \vec A_c (\vec k) - \vec A_v (\vec k) - \nabla_{\vec k} \arg [\nu_\theta (\vec k)]. 
\ee 
For our symmetry and stacking analysis below, it will be useful to re-express this conventional form of the shift vector in terms of a Wilson line \cite{ShiftVector, PhotonDrag}:
\be 
\vec r (\theta, \vec k) =  \lim_{\vec q \to 0} \nabla_{\vec q} \arg [\mathcal W (\theta, \vec k, \vec q)],
\ee
where $\theta$ is the electric field polarisation angle of the incident light with respect to the $x$, $\vec k$ is the wavevector measured from the $\Gamma$ point, and 
\begin{align}
\mathcal W (\theta, \vec k, \vec q) = & \la u_v (\vec k ) | u_v (\vec k + \vec q ) \ra \la u_v (\vec k + \vec q ) | \nu_\theta | u_c (\vec k + \vec q ) \ra \la u_c (\vec k + \vec q ) | u_c (\vec k ) \ra. 
\end{align}
Here $| u_{c(v)} (\vec k) \ra$ is the Bloch wavefunction of the conduction (valence) band. The velocity matrix $\nu_\theta = \hat{\nu} \cdot \hat{\vec e}_\theta$ is a function of the polarisation direction $\hat{\vec e}_\theta$. For linearly polarised light, we have $\nu_\theta = \nu_x \cos \theta + \nu_y \sin \theta$. We note that though the Wilson line $\mathcal W (\theta, \vec k, \vec q)$ depends on the gauge choice of the wavefunction, the gradient of its phase $\nabla_{\vec q}  \arg [\mathcal W (\theta, \vec k, \vec q)]$ and the shift vector are gauge invariant. In the following, we examine the properties of $\la u_n (\vec k)| u_m (\vec q) \ra$ and $\la u_n (\vec k)| \nu_\theta | u_m (\vec q) \ra$ in different vdW materials and homostructures and the atomic configuration dependence of the shift vector and SPC. In particular, we illustrate the configuration dependent SPC in Bernal stacked bilayer graphene (BLG), graphene on hexagonal boron nitride (G/hBN), monolayer transition metal dicalcogenide (TMD) and 2H stacked bilayer TMD. 

\subsection{Shift Vector Configuration Dependence for Bernal Stacked BLG}

Bernal stacked BLG possesses a three-fold rotational symmetry $C^z_3$ and mirror symmetry about the armchair direction, as shown in Fig.~1 in the main text (e.g., the $y$-axis in Fig.~1). Applying an interlayer electric potential difference $\Delta$ breaks the inversion symmetry of the system. As we see below, this gives rise to a nonzero SPC. 

Two stacking configurations are possible in Bernal stacked BLG: AB stacking whereby the A site of the top layer is directly on top of the B site of the bottom layer and BA stacking whereby the B site of the top layer is directly on top of the A site of the bottom layer. Here we describe AB/BA stacked BLG with the real space Hamiltonian $\mathcal H^{(\eta)} (\Delta, \vec r)$, where $\eta = {\rm AB, BA}$ denotes the stacking configuration and $\Delta$ denotes the interlayer potential difference. In the following, we examine the symmetry constraints of the shift vector $\vec r^{(\eta)} (\Delta, \theta, \vec k)$ and the stacking and configuration dependence of the SPC.

\subsubsection{Time Reversal Symmetry}
BLG exhibits spin degeneracy and can be considered an effectively spinless system. In the presence of time reversal symmetry $\mathcal T$, the Bloch Hamiltonian $H^{(\eta)} (\Delta, \vec k) = e^{-i \vec k \cdot \vec r } \mathcal H^{(\eta)} (\Delta, \vec r) e^{i \vec k \cdot \vec r } $ satisfies $\mathcal T H^{(\eta)} (\Delta, \vec k) \mathcal T^{-1} = H^{(\eta)} (\Delta, - \vec k)$. The Bloch wavefunction of band $n$, $| u_n^{(\eta)} (\Delta, \vec k ) \ra$ is defined such that $H^{(\eta)} (\Delta, \vec k) | u_n^{(\eta)} (\Delta, \vec k ) \ra = \epsilon_n^{(\eta)} (\Delta, \vec k) | u_n^{(\eta)} (\Delta, \vec k ) \ra$, where $\epsilon_n^{(\eta)} (\Delta, \vec k)$ is the energy eigenvalue of band $n$.
By considering $\mathcal T H^{(\eta)} (\Delta, \vec k)  | u_n^{(\eta)} (\Delta, \vec k ) \ra$, we find that the Bloch wavefunction transforms as:
\be 
\epsilon_n^{(\eta)} (\Delta, \vec k) = \epsilon_n^{(\eta)} (\Delta, - \vec k), \quad \mathcal T |u_n^{(\eta)} (\Delta, \vec k) \ra = | u_n^{(\eta)} (\Delta, -\vec k) \ra^*.  
\ee
Thus, for any wavevectors $\vec k_1$ and $\vec k_2$ measured from the centre of the Brillouin zone, we have 
\be
\la u_m^{(\eta)} (\Delta, \vec k_1) | u_n^{(\eta)} (\Delta, \vec k_2) \ra = \la u_m^{(\eta)} (\Delta, \vec k_1) | \mathcal T^{-1} \mathcal T | u_n^{(\eta)} (\Delta, \vec k_2) \ra = \la u_m^{(\eta)} (\Delta, - \vec k_1) | u_n^{(\eta)} (\Delta, -\vec k_2) \ra^*. 
\ee
The velocity operator is odd under time reversal: $\mathcal T \hat \nu^{(\eta)} (\Delta) \mathcal T^{-1} = - \hat \nu^{(\eta)} (\Delta)$. Thus, the velocity matrix element satisfies 
\begin{align}
\la u_m^{(\eta)} (\Delta, \vec k_1) | \nu_\theta^{(\eta)} (\Delta) | u_n^{(\eta)} (\Delta, \vec k_2) \ra &= \la u_m^{(\eta)} (\Delta, \vec k_1) | \mathcal T^{-1} \mathcal T \nu_\theta^{(\eta)} (\Delta) \mathcal T^{-1} \mathcal T | u_n^{(\eta)} (\Delta, \vec k_2) \ra \nonumber \\
&= - \la u_m^{(\eta)} (\Delta, -\vec k_1) | \nu_\theta^{(\eta)} (\Delta) | u_n^{(\eta)} (\Delta, -\vec k_2) \ra^*. 
\end{align}
Similarly, we obtain the symmetry constraint for the Wilson line: 
\be 
\mathcal W^{(\eta)} (\Delta, \theta, \vec k, \vec q) = - [\mathcal W^{(\eta)} (\Delta, \theta, -\vec k, - \vec q)]^*, \quad \arg [ \mathcal W^{(\eta)} (\Delta, \theta, \vec k, \vec q)] = - \arg [ \mathcal W^{(\eta)} (\Delta, \theta, - \vec k, -\vec q)] + \pi.
\ee
As a result, the shift vector in a time-reversal invariant system is even in $k$-space: 
\be\label{Seq:T} 
\vec r^{(\eta)} (\Delta, \theta, \vec k) =  \vec r^{(\eta)} (\Delta, \theta, -\vec k). 
\ee

\subsubsection{Inversion Operation and Interlayer Potential Dependence}

In AB/BA stacked bilayer BLG, the interlayer potential difference $\Delta$ breaks the inversion symmetry. Here we show that breaking of the inversion symmetry is necessary to induce a finite SPC. Furthermore, switching the direction of the interlayer potential flips the sign of the shift vector -- as a result, SPC flows in the opposite direction. 

To see this, we observe that upon spatial inversion, the atomic configuration of the BLG remains unchanged but the interlayer potential difference $\Delta$ flips sign. The real-space Hamiltonian satisfies $\mathcal I \mathcal H^{(\eta)} (\Delta, \vec r) \mathcal I^{-1}= \mathcal H^{(\eta)} (\Delta, - \vec r) = \mathcal H^{(\eta)} (-\Delta, \vec r)$. Thus, under inversion, the Bloch Hamiltonian transforms as $ \mathcal I H^{(\eta)} (\Delta, \vec k) \mathcal I^{-1} = H^{(\eta)} (-\Delta, - \vec k)$. By considering $\mathcal I H^{(\eta)} (\vec k ) | u_n^{(\eta)} (\vec k ) \ra$, one arrives  at 
\be
\epsilon_n^{(\eta)} (\Delta, \vec k) = \epsilon_n^{(\eta)} (-\Delta, -\vec k), \quad \mathcal I | u_n^{(\eta)} (\Delta, \vec k) \ra = | u_n^{(\eta)} (-\Delta, -\vec k) \ra
\ee
This gives 
\be
\la u_m^{(\eta)} (\Delta, \vec k_1) | u_n^{(\eta)} (\Delta, \vec k_2) \ra = \la u_m^{(\eta)} (\Delta, \vec k_1) | \mathcal I^{-1} \mathcal I | u_n^{(\eta)} (\Delta, \vec k_2) \ra = \la u_m^{(\eta)} ( -\Delta, -\vec k_1) | u_n^{(\eta)} ( -\Delta, -\vec k_2) \ra. 
\ee

Upon inversion, velocity operator transforms as $\mathcal I \hat \nu^{(\eta)} (\Delta) \mathcal I^{-1} = - \hat \nu^{(\eta)} (-\Delta)$. Thus for electric field polarisation $\theta$, we have 
\begin{align}
\la u_m^{(\eta)} (\Delta, \vec k_1) | \nu_\theta (\Delta) | u_n^{(\eta)} (\Delta, \vec k_2) \ra &= \la u_m^{(\eta)} (\Delta, \vec k_1) | \mathcal I^{-1} \mathcal I \nu_\theta^{(\eta)} (\Delta) \mathcal I^{-1} \mathcal I | u_n^{(\eta)} (\Delta, \vec k_2) \ra \nonumber \\
&= - \la u_m^{(\eta)} ( -\Delta, -\vec k_1) |\nu_\theta^{(\eta)} (-\Delta) | u_n^{(\eta)} ( -\Delta, -\vec k_2) \ra.
\end{align}
Using these relations above, the Wilson line satisfies 
\be 
\mathcal W^{(\eta)} (\Delta, \theta, \vec k, \vec q) = -\mathcal W^{(\eta)} (-\Delta, \theta, -\vec k, -\vec q),\quad \arg\left[\mathcal W^{(\eta)} (\Delta, \theta, \vec k, \vec q) \right] = \arg \left[ \mathcal W^{(\eta)} (-\Delta, \theta, -\vec k, -\vec q) \right] +\pi,
\ee
and the shift vector satisfies 
\be 
\vec r^{(\eta)} (\Delta, \theta, \vec k) = - \vec r^{(\eta)} (-\Delta, \theta, - \vec k).
\ee
Importantly, when $\Delta = 0$, the above relation demands that the shift vector is odd in the $k$-space. Since $\rho^{(\eta)} (\Delta, \vec k)$ and $| \nu_\theta^{(\eta)} (\Delta)|^2$ are both even, upon integrating the weighted shift vector in the Brillouin zone [see Eq.~(2) of the main text], we obtain the well-known vanishing of SPC in an inversion symmetric system. Thus, inversion symmetry has to be broken to obtain finite SPC. 

We note that by further applying time reversal symmetry (see above section) where $\vec r^{(\eta)}$ is even in the $k$-space [Eq.~(\ref{Seq:T})], we arrive at the dependence of shift vector on the sign of interlayer potential: 
\be\label{Seq:rDelta}
\vec r^{(\eta)} (\Delta,\theta, \vec k) = - \vec r^{(\eta)} (-\Delta, \theta, \vec k).
\ee
The shift vector switches sign when the interlayer potential difference is reversed. We see in Fig.~1b in the main text that this is manifested in opposite SPC for $\Delta \to - \Delta$.

\subsubsection{Mirror Symmetry}
Bernal stacked BLG exhibits mirror symmetry about the armchair directions. In the coordinate system shown in Fig.~1 in the main text, one of the mirror reflection axis is along the $y$ direction;  the system is invariant under the mirror operation $\mathcal M_y: (x, y,z) \to ( - x , y, z)$. The real space Hamiltonian obeys $\mathcal M_y \mathcal H^{(\eta)} (\Delta, \vec r) \mathcal M_y^{-1} = \mathcal H^{(\eta)} (\Delta, \vec r)$. The Bloch Hamiltonian thus satisfies $\mathcal M_y H^{(\eta)} (\Delta, \vec k) \mathcal M_y^{-1} = H^{(\eta)} (\Delta, \mathcal M_y \vec k)$ and we have 
\be 
\epsilon_n^{(\eta)} (\Delta, \vec k) = \epsilon_n^{(\eta)} (\Delta, \mathcal M_y \vec k), \quad \mathcal M_y |u_n^{(\eta)} (\Delta, \vec k) = | u_n^{(\eta)} (\Delta, \mathcal  M_y \vec k) \ra.  
\ee
The above relation gives
\be 
\la u_m^{(\eta)} (\Delta, \vec k_1) | u_n^{(\eta)} (\Delta, \vec k_2) \ra = \la u_m^{(\eta)} (\Delta, \vec k_1) | \mathcal M_y^{-1} \mathcal M_y | u_n^{(\eta)} (\Delta, \vec k_2) \ra = \la u_m^{(\eta)} (\Delta, \mathcal M_y \vec k_1) | u_n^{(\eta)} (\Delta, \mathcal M_y \vec k_2) \ra. 
\ee

Under $\mathcal M_y$, the $x$ component of the velocity operator switches sign $\mathcal M_y \nu_x^{(\eta)} (\Delta) \mathcal M_y^{-1} = - \nu_x^{(\eta)} (\Delta)$ while the $y$ component remains invariant $\mathcal M_y \nu_y^{(\eta)} (\Delta) \mathcal M_y^{-1} = \nu_y^{(\eta)} (\Delta)$, thus we have $\mathcal M_y \nu_\theta^{(\eta)} (\Delta) \mathcal M_y^{-1} =  \nu_{\pi - \theta}^{(\eta)} (\Delta)$. This gives 
\begin{align}
\la u_m^{(\eta)}  (\Delta, \vec k_1) | \nu_\theta^{(\eta)} (\Delta) | u_n^{(\eta)}  (\Delta, \vec k_2) \ra &= \la u_m^{(\eta)}  (\Delta, \vec k_1) | \mathcal M_y^{-1} \mathcal M_y \nu_\theta^{(\eta)} (\Delta) \mathcal M_y^{-1} \mathcal M_y | u_n^{(\eta)}  (\Delta, \vec k_2) \ra \nonumber \\
&= \la u_m^{(\eta)}  (\Delta, \mathcal M_y \vec k_1) | \nu_{\pi - \theta}^{(\eta)} (\Delta) | u_n^{(\eta)}  (\Delta, \mathcal M_y \vec k_2) \ra. 
\end{align}
Using these relations above, the Wilson line satisfies 
\be
\mathcal W^{(\eta)}  (\Delta, \theta, \vec k, \vec q) = \mathcal W^{(\eta)}  (\Delta, \pi - \theta, \mathcal M_y  \vec k, \mathcal M_y \vec q), \quad \arg [ \mathcal W^{(\eta)} (\Delta, \theta, \vec k,  \vec q)] = \arg [ \mathcal W^{(\eta)} (\Delta, \pi- \theta, \mathcal M_y \vec k, \mathcal M_y \vec q)]  , 
\ee
and shift vector thus satisfies 
\be\label{Seq:My_intermediate}
r_x^{(\eta)}  (\Delta, \theta, \vec k) = - r_x^{(\eta)}  (\Delta, \pi - \theta, \mathcal M_y \vec k) , \quad r_y^{(\eta)}  (\Delta, \theta, \vec k) = r_y^{(\eta)}  (\Delta, \pi - \theta, \mathcal M_y \vec k). 
\ee

We note, parenthetically, for a given ac electric field (light irradiation), the polarisations along $\hat{\vec e}$ and $-\hat{\vec e}$ directions are equivalent and yield the same shift vector. To see this, we observe that $\nu_\theta^{(\eta)} (\Delta) = - \nu_{\theta+\pi}^{(\eta)} (\Delta)$ and
\be
\mathcal W^{(\eta)} (\Delta, \theta, \vec k, \vec q) = -\mathcal W^{(\eta)} (\Delta, \theta+\pi, \vec k, \vec q), \quad \arg [ \mathcal W^{(\eta)} (\Delta, \theta, \vec k, \vec q)] = \arg [ \mathcal W^{(\eta)} (\Delta, \theta+\pi, \vec k, \vec q)] + \pi. 
\ee
Since the shift vector is the gradient of the argument of the Wilson line, an additional $\pi$ phase shift for $ \mathcal W^{(\eta)} (\Delta, \theta, \vec k, \vec q)$ does not affect the shift vector
\be\label{Seq:rtheta_pi}
\vec r^{(\eta)} (\Delta, \theta, \vec k) = \vec r^{(\eta)} (\Delta, \theta+\pi, \vec k).
\ee
Thus, Eq. (\ref{Seq:My_intermediate}) can be rewritten as 
\be\label{Seq:My}
r_x^{(\eta)}  (\Delta, \theta, \vec k) = - r_x^{(\eta)}  (\Delta, - \theta, \mathcal M_y \vec k) , \quad r_y^{(\eta)}  (\Delta, \theta, \vec k) = r_y^{(\eta)}  ( \Delta, - \theta, \mathcal M_y \vec k). 
\ee

Furthermore, combining Eq. (\ref{Seq:T}) and (\ref{Seq:My}), the composition of time reversal $\mathcal T$ and mirror symmetry $\mathcal M_y$ yields 
\be\label{Seq:TMy}
r_x^{(\eta)}  (\Delta, \theta, \vec k) = - r_x^{(\eta)}  (\Delta, - \theta, k_x, -k_y) , \quad r_y^{(\eta)}  (\Delta, \theta, \vec k) = r_y^{(\eta)}  (\Delta,  - \theta, k_x, -k_y). 
\ee
Eq.~(\ref{Seq:TMy}) gives the symmetry constraints for the shift vector, which is manifested as the constraints for the direction of the SPC when the electric field is polarised along high-symmetry axes. For example, when the electric field polarisation is normal to the mirror plane (i..e $\theta = 0, \pi$), the shift vector satisfies 
\be\label{Seq:xpol_TMy}
r_x^{(\eta)}  (\Delta, 0, \vec k) = - r_x^{(\eta)}  (\Delta, 0, k_x, -k_y) , \quad r_y^{(\eta)}  (\Delta, 0, \vec k) = r_y^{(\eta)}  ( \Delta, 0, k_x, -k_y). 
\ee
Since both $\rho (\vec k)$ and $|\nu_\theta^{(\eta)} (\Delta)|^2$ are even under $k_y \to -k_y$, the symmetry constraint in Eq.~(\ref{Seq:xpol_TMy}) ensures that upon integration in the $k$-space [Eq.~(2) in the main text], the $x$ component of the SPC vanishes while the $y$ component is nonzero. Thus, for linear polarisation normal to the mirror plane, SPC is completely transverse. 

Similarly, when the electric field is polarised along the mirror reflection axis (i.e. $\theta = \pm \pi/2$), Eq.~(\ref{Seq:TMy}) reduces to 
\be\label{Seq:ypol_TMy}
r_x^{(\eta)}  (\Delta, \pi/2, \vec k) = - r_x^{(\eta)}  (\Delta, \pi/2, k_x, -k_y) , \quad r_y^{(\eta)}  (\Delta, \pi/2, \vec k) = r_y^{(\eta)}  (\Delta,  \pi/2, k_x, -k_y). 
\ee
Since $r_x^{(\eta)}  (\Delta, \pi/2, \vec k)$ is odd in $k_y$ while $r_y^{(\eta)}  (\Delta, \pi/2, \vec k)$ is even, Eq.~(\ref{Seq:ypol_TMy}) implies that SPC is along the $y$ direction. Thus, for electric field polarisation parallel to the mirror plane, we obtain completely longitudinal SPC.

\subsubsection{In-plane Three-fold Rotational Symmetry}
The BLG lattice is invariant under in-plane three-fold rotational symmetry $C^z_3$ such that $C^z_3 \mathcal H^{(\eta)} (\Delta, \vec r) (C^z_3)^{-1} = \mathcal H^{(\eta)} (\Delta, \vec r)$. The Bloch Hamiltonian thus obeys the relation $C^z_3 H^{(\eta)} (\Delta, \vec k) (C^z_3)^{-1} =  H^{(\eta)} (\Delta, C^z_3 \vec k)$. Here, $C^z_3 \vec k$ is defined by the rotation matrix 
\be
C^z_3 \begin{pmatrix} k_x \\ k_y \end{pmatrix} = \begin{pmatrix} 
\cos \vartheta & \sin \vartheta \\
-\sin \vartheta & \cos \vartheta \end{pmatrix} \begin{pmatrix} k_x \\ k_y \end{pmatrix},
\ee
where $\vartheta = 2\pi/3$. 

The $C^z_3$ amd $\mathcal M_y$ symmetries imply that the system preserves reflection symmetry about other two axes with angle $\pm \pi/6$ with respect from the $x$-axis. Thus, for electric fields polarised parallel or normal to those two high-symmetry axes, the shift vector also acts as a pseudovector.

\subsubsection{Stacking Dependence}
Now we show that the shift vector switches sign when the stacking configuration is switched from AB to BA. AB/BA stacked BLG are related by flipping the sample about an axes along the armchair direction. For concreteness, we concentrate on the $y$-axis, so that this flip can be denoted $C^y_2: (x, y, z ) \to (-x, y, -z)$ as the rotation about $y$-axis by $\pi$. We note that $C^y_2$ not only flips the sample, but also interchanges the layer indices. As a result, the interlayer potential between the bottom and the top layer $\Delta$ switches sign. The real-space Hamiltonians of AB and BA stacked BLG satisfy $C^y_2 \mathcal H^{\rm (AB)} (\Delta, \vec r) (C^y_2)^{-1} = \mathcal H^{\rm (AB)} (\Delta, C^y_2 \vec r) = \mathcal H^{\rm (BA)} (-\Delta, \vec r)$. The Bloch Hamiltonians are thus related by $C^y_2 H^{\rm (AB)} (\Delta, \vec k) (C^y_2)^{-1} = H^{\rm (BA)}(- \Delta, C^y_2 \vec k) $. Here, we consider BLG as a 2D system with $\vec k = (k_x, k_y)$ and $C^y_2 (k_x, k_y) = (- k_x, k_y) = \mathcal M_y (k_x, k_y)$. We note that the dispersion is independent of stacking configuration and interlayer potential direction $\epsilon^{\rm (AB)}_n (\Delta, \vec k) = \epsilon^{\rm (BA)}_n (-\Delta, \vec k) $ and $C^y_2$ satisfies $(C^y_2)^{-1} = C^y_2 = (C^y_2)^\dag$. 
It follows that the wavefunctions are related by 
\be
\epsilon^{\rm (AB)}_{n} (\Delta, \vec k) = \epsilon_{ n}^{\rm (BA)} (-\Delta, C^y_2 \vec k), \quad C^y_2 |u_{ n}^{\rm (AB)} (\Delta, \vec k) = | u_{n}^{\rm (BA)} (-\Delta, C^y_2 \vec k) \ra.  
\ee
Thus we have 
\be 
\la u_{m}^{\rm (AB)} (\Delta, \vec k_1) | u_{n}^{\rm (AB)} (\Delta, \vec k_2) \ra = \la u_{m}^{\rm (AB)} (\Delta, \vec k_1) | (C^y_2)^{-1}C^y_2 | u_{n}^{\rm (AB)} (\Delta, \vec k_2) \ra = \la u_{m}^{\rm (BA)} (-\Delta, C^y_2 \vec k_1) | u_{n}^{\rm (BA)} (-\Delta, C^y_2 \vec k_2) \ra. 
\ee

The velocity operators transform as $C^y_2 \nu_{ x}^{\rm (AB)} (\Delta) (C^y_2 )^{-1} = -\nu_{x}^{\rm (BA)} (-\Delta) $ and $C^y_2 \nu_{y}^{\rm (AB)} (\Delta) (C^y_2 )^{-1} = \nu_{y}^{\rm (BA)} (-\Delta)$. Thus, for a linear polarisation angle $\theta$, we have $C^y_2 \nu_{\theta}^{\rm (AB)} (\Delta) (C^y_2 )^{-1} = \nu_{\pi -\theta}^{\rm (BA)} (-\Delta)$. This gives 
\begin{align}
\la u_{m}^{\rm (AB)} (\Delta, \vec k_1) | \nu_{\theta}^{\rm (AB)} (\Delta) | u_{n}^{\rm (AB)} (\Delta, \vec k_2) \ra &= \la u_{m}^{\rm (AB)} (\Delta, \vec k_1) | (C^y_2)^{-1} C^y_2 \nu_{ \theta}^{\rm (AB)} (\Delta)  (C^y_2 )^{-1} C^y_2 | u_{n}^{\rm (AB)} (\Delta, \vec k_2) \ra \nonumber \\
&= \la u_{m}^{\rm (BA)} (-\Delta, C^y_2 \vec k_1) | \nu_{\pi - \theta}^{\rm (BA)} (-\Delta) | u^{(\rm BA)}_{n} (-\Delta, C^y_2 \vec k_2) \ra. 
\end{align}
We can perform the similar analysis on the Wilson line and the shift vector: 
\be
\mathcal W^{\rm (AB)} (\Delta, \theta, \vec k, \vec q) = \mathcal W^{\rm (BA)} (-\Delta, \pi- \theta, C^y_2  \vec k, C^y_2 \vec q), \quad \arg [ \mathcal W^{\rm (AB)} (\Delta, \theta, \vec k,  \vec q)] =  \arg [ \mathcal W^{\rm (BA)} (-\Delta, \pi- \theta, C^y_2 \vec k, C^y_2 \vec q)], 
\ee
The shift vector can be calculated by taking the derivatives of $\arg [ \mathcal W^{(\eta)}]$ in $k$-space. Noting the identity in Eq.~(\ref{Seq:rtheta_pi}) that the shift vector is invariant for $\theta \to \theta + \pi$, we arrive at  
\be\label{eq:r_Cx2}
r_{x}^{\rm (AB)} (\Delta, \theta, \vec k) = - r_{x}^{\rm (BA)} (-\Delta, - \theta, C^y_2 \vec k) , \quad r_{y}^{\rm (AB)} (\Delta, \theta, \vec k) = r_{y}^{\rm (BA)} ( -\Delta, - \theta, C^y_2 \vec k). 
\ee
Noting that $\mathcal M_y C^y_2 \vec k = \vec k$, and combining Eq. (\ref{Seq:My}) and (\ref{eq:r_Cx2}), we have 
\be
\vec r^{\rm (AB)} (\Delta, \theta, \vec k) =  \vec r^{\rm (BA)} (-\Delta, \theta, \vec k).
\ee
Finally, recalling that the shift vector switches sign for $\Delta \to -\Delta$ [Eq.~(\ref{Seq:rDelta})], thus we have 
\be
\vec r^{\rm (AB)} (\Delta, \theta, \vec k) = - \vec r^{\rm (BA)} (\Delta, \theta, \vec k).
\ee
Since the dispersion and $|\nu_\theta^{(\eta)}|^2$ are stacking independent, it follows that the shift currents flow in opposite directions in the two stacking configurations, as discussed in the main text.

\subsection{Shift Vector Configuration Dependence for Staggered Sublattice Potential, e.g. G/hBN}

We now examine the symmetry constraints of the shift vector in a gapped Dirac material with staggered sublattice potential difference described by the Bloch Hamiltonian $H (\delta, \vec k)$, where $\delta$ is the sublattice potential difference and $\vec k$ is the wavevector measured from the $\Gamma$ point. In this section, we focus on the spinless fermions, which can be realised in a commensurate stacked graphene-hexagonal boron nitride (G/hBN) system, where the sign of $\delta$ depends on the alignment between the graphene and hBN layers. Such commensurate stacking have been achieved recently in G/hBN as evidenced by substantial gap opening at the charge neutrality point. For simplicity, we will consider the case whereby the carbon atoms in graphene are directly on top of the boron and nitrogen atoms in hBN (e.g., found within a single commensurate domain). 
In-plane rotation of the hBN layer by $\pi$ (keeping the graphene layer fixed) leads to the interchange of the electric potential at the A and B site of graphene, thus reversing the sign of $\delta$. In the following, we will show that the shift vector and SPC depend on the atomic alignment and light polarisation. For consistency of notation for the various vdW materials and systems considered here, we will fix the orientation so that one of the armchair directions is aligned along the $y$-direction (similar to that discussed for BLG above).

\subsubsection{Time Reversal Symmetry}

The system is invariant under time reversal operation: $\mathcal T H (\delta, \vec k) \mathcal T^{-1} = H(\delta, - \vec k)$. Thus the dispersion and the Bloch wavefunctions transform under $\mathcal T $ as 
\be
\epsilon_n (\delta, \vec k) = \epsilon_n (\delta, - \vec k), \quad \mathcal T |u_n (\delta, \vec k) \ra = | u_n (\delta, -\vec k) \ra^*.  
\ee
The above gives us
\be
\la u_m (\delta, \vec k_1) | u_n (\delta, \vec k_2) \ra = \la u_m (\delta, \vec k_1) | \mathcal T^{-1} \mathcal T | u_n (\delta, \vec k_2) \ra = \la u_m (\delta,  - \vec k_1) | u_n (\delta,  -\vec k_2) \ra^*. 
\ee
The velocity operator transforms as $\mathcal T \hat \nu(\delta) \mathcal T^{-1} = - \hat \nu (\delta)$. Thus, for a given polarisation angle $\theta$, the velocity matrix element satisfies 
\begin{align}
\la u_m (\delta, \vec k_1) | \nu_\theta (\delta) | u_n (\delta, \vec k_2) \ra &= \la u_m (\delta, \vec k_1) | \mathcal T^{-1} \mathcal T [\nu_x (\delta) \cos \theta + \nu_y (\delta) \sin \theta] \mathcal T^{-1} \mathcal T | u_n (\delta, \vec k_2) \ra \nonumber \\
&= - \la u_m (\delta, -\vec k_1) | \nu_\theta (\delta) | u_n (\delta, -\vec k_2) \ra^*. 
\end{align}
We obtain the symmetry constraint for the Wilson line: 
\be 
\mathcal W (\delta, \theta, \vec k, \vec q) = - [\mathcal W (\delta, \theta, -\vec k, - \vec q)]^*, \quad \arg [ \mathcal W (\delta, \theta, \vec k, \vec q)] = - \arg [ \mathcal W (\delta, \theta, - \vec k, -\vec q)] + \pi.
\ee
As a result, the shift vector in a time-reversal invariant system is even in $k$-space: 
\be\label{Seq:GhBN_T} 
\vec r (\delta, \theta, \vec k) =  \vec r (\delta, \theta, -\vec k). 
\ee

\subsubsection{Inversion}

For $\delta \neq 0$, the system breaks inversion symmetry. Under inversion operation, the real space Hamiltonian satisfies $\mathcal I \mathcal H (\delta, \vec r) \mathcal I^{-1} = \mathcal H (\delta, -\vec r) = \mathcal H (-\delta, \vec r) $, i.e. inversion switches the staggered potential. As a result, the Bloch Hamiltonian transforms as $\mathcal I H (\delta, \vec k) \mathcal I^{-1} = H (-\delta, -\vec k)$, and the dispersion and Bloch wavefunction satisfy: 
\be
\epsilon_n (\delta, \vec k) = \epsilon_n (-\delta, - \vec k), \quad \mathcal I |u_n (\delta, \vec k) \ra = | u_n (-\delta, -\vec k) \ra.  
\ee
Thus we have 
\be
\la u_m (\delta, \vec k_1) | u_n (\delta, \vec k_2) \ra = \la u_m (\delta, \vec k_1) | \mathcal I^{-1} \mathcal I | u_n (\delta, \vec k_2) \ra = \la u_m (-\delta,  - \vec k_1) | u_n (-\delta,  -\vec k_2) \ra. 
\ee

On the other hand, the velocity operator transforms as $\mathcal I \hat \nu(\delta) \mathcal I^{-1} = - \hat \nu (-\delta)$. This yields
\begin{align}
\la u_m (\delta, \vec k_1) | \nu_\theta (\delta) | u_n (\delta, \vec k_2) \ra &= \la u_m (\delta, \vec k_1) | \mathcal I^{-1} \mathcal I \nu_\theta (\delta) \mathcal I^{-1} \mathcal I | u_n (\delta, \vec k_2) \ra \nonumber \\
&= - \la u_m (-\delta, -\vec k_1) | \nu_\theta (-\delta) | u_n (-\delta, -\vec k_2) \ra. 
\end{align}
We obtain the symmetry constraint for the Wilson line: 
\be 
\mathcal W (\delta, \theta, \vec k, \vec q) = - \mathcal W (-\delta, \theta, -\vec k, - \vec q), \quad \arg [ \mathcal W (\delta, \theta, \vec k, \vec q)] =  \arg [ \mathcal W (-\delta, \theta, - \vec k, -\vec q)] + \pi.
\ee
The shift vector obeys the following relation:
\be\label{Seq:GhBN_I} 
\vec r (\delta, \theta, \vec k) = - \vec r (-\delta, \theta, -\vec k). 
\ee

Furthermore, we note that under time reversal symmetry, the shift vector is even in $k$-space. Eq.~(\ref{Seq:GhBN_T}) and~(\ref{Seq:GhBN_I}) demand that the shift vector switches sign when the sublattice potential difference is switched:
\be\label{Seq:GhBN_delta} 
\vec r (\delta, \theta, \vec k) = - \vec r (-\delta, \theta, \vec k). 
\ee
As a result, the direction of SPC is expected to be reversed when the sublattice potential difference is reversed. As we discussed, in G/hBN, this can be achieved by different alignment of hBN below the graphene layer, for example, by in-plane rotation of hBN by $\pi$. Thus SPC serves a a tool to determine the stacking alignment in G/hBN.

\subsubsection{Mirror Symmetry}

For consistency of notation for the various vdW materials and systems considered here,
we will fix the orientation so that one of the armchair directions is aligned along the $y$-direction (similar to that discussed for BLG above). Mirror symmetry about the $y$-axis ensures: 
$\mathcal M_y \mathcal H (\delta, \vec r) \mathcal M_y^{-1} = \mathcal H(\delta, \vec r)$ and $\mathcal M_y H (\delta, \vec k) \mathcal M_y^{-1} = H (\delta, \mathcal M_y \vec k)$. Thus we have 
\be
\epsilon_n (\delta, \vec k) = \epsilon_n (\delta, \mathcal M_y \vec k), \quad \mathcal M_y |u_n (\delta, \vec k) \ra = | u_n (\delta, \mathcal M_y \vec k) \ra,  
\ee
and
\be
\la u_m (\delta, \vec k_1) | u_n (\delta, \vec k_2) \ra = \la u_m (\delta, \vec k_1) | \mathcal M_y^{-1} \mathcal M_y | u_n (\delta, \vec k_2) \ra = \la u_m (\delta,  \mathcal M_y \vec k_1) | u_n (\delta,  \mathcal M_y \vec k_2) \ra. 
\ee

The velocity operators transform as $\mathcal M_y \nu_x (\delta)  \mathcal M_y^{-1} = - \nu_x (\delta) $ and $\mathcal M_y \nu_y (\delta)  \mathcal M_y^{-1} = \nu_y (\delta) $, thus we have $\mathcal M_y \nu_\theta^{(\eta)}  \mathcal M_y^{-1} =  \nu_{\pi - \theta} (\delta) $. This gives 
\begin{align}
\la u_m  (\delta, \vec k_1) | \nu_\theta (\delta)  | u_n (\delta, \vec k_2) \ra &= \la u_m  (\delta, \vec k_1) | \mathcal M_y^{-1} \mathcal M_y \nu_\theta (\delta)  \mathcal M_y^{-1} \mathcal M_y | u_n  (\delta, \vec k_2) \ra \nonumber \\
&= \la u_m (\delta, \mathcal M_y \vec k_1) | \nu_{\pi - \theta} (\delta)  | u_n (\delta, \mathcal M_y \vec k_2) \ra. 
\end{align}
The Wilson line satisfies 
\be
\mathcal W (\delta, \theta, \vec k, \vec q) = \mathcal W (\delta, \pi - \theta, \mathcal M_y  \vec k, \mathcal M_y \vec q), \quad \arg [ \mathcal W (\delta, \theta, \vec k,  \vec q)] = \arg [ \mathcal W (\delta, \pi- \theta, \mathcal M_y \vec k, \mathcal M_y \vec q)]  , 
\ee
and shift vector thus satisfies 
\be\label{Seq:hBN_My_intermediate}
r_x  (\delta, \theta, \vec k) = - r_x  (\delta, \pi - \theta, \mathcal M_y \vec k) , \quad r_y  (\delta, \theta, \vec k) = r_y  (\delta, \pi - \theta, \mathcal M_y \vec k). 
\ee

We note that the electric field polarisation along $\hat{\vec e}_\theta$ is equivalent to the polarisation along $- \hat{\vec e}_\theta = \hat{\vec e}_{\pi+\theta}$. Thus Eq.~(\ref{Seq:hBN_My_intermediate}) can be rewritten as 
\be\label{Seq:hBN_My}
r_x  (\delta, \theta, \vec k) = - r_x  (\delta, - \theta, \mathcal M_y \vec k) , \quad r_y  (\delta, \theta, \vec k) = r_y  (\delta, - \theta, \mathcal M_y \vec k). 
\ee

Additionally, following the similar analysis in Eq.~(\ref{Seq:TMy}), we combine the constraints of time reversal symmetry [Eq.~(\ref{Seq:GhBN_T})] and mirror symmetry [Eq.~(\ref{Seq:hBN_My})] to obtain   
\be\label{Seq:hBN_TMy}
r_x  (\delta, \theta, \vec k) = - r_x  (\delta, - \theta, k_x, - k_y) , \quad r_y  (\delta, \theta, \vec k) = r_y  (\delta, - \theta, k_x, -k_y). 
\ee

Similar with the scenario of Bernal stacked BLG, the shift vector in Eq.~(\ref{Seq:hBN_TMy}) also exhibits pseudovector property when the electric field polarisation is either perpendicular to ($\theta = 0, \pi$) or parallel with ($\theta = \pm \pi/2$) with mirror plane ($y$-axis). At these polarisations, $r_x (\delta, \theta, \vec k)$ flips sign for $k_y \to - k_y$ while $r_y (\delta, \theta, \vec k)$ remains invariant. Thus, integration over the $k$-space in Eq.~(2) in the main text leads to transverse (longitudinal) SPC for linear polarisations perpendicular to (parallel with) the mirror reflection axis. 

Furthermore, the material under consideration is also invariant under three-fold in-plane rotational symmetry $C^z_3$ like BLG. Thus the system possesses three mirror reflection axes with angle $\pi/2, \pm \pi /6$. Similar argument can be applied to the shift vector and shift current after rotating the system by $2 \pi/3$ about the $z$ axis, and we expect transverse (longitudinal) SPC for polarisations perpendicular to (parallel with) any of the mirror reflection axes.

\subsection{Shift Vector Configuration Dependence for Monolayer TMDs}

Now we consider the configuration dependence of shift vector in monolayer TMDs, which possess hexagonal lattice structure with A and B lattice site hosting different atoms. Thus they have sublattice potential difference $\delta$. Further, large Ising spin-orbit coupling yields spin-valley locked states. In the following, we show that the spin SPC in these materials depends on the $\delta$ and the incident light polarisation.

\subsubsection{Time Reversal Symmetry}
Monolayer TMDs are invariant under time reversal symmetry $\mathcal T H (\delta, \vec k) \mathcal T^{-1} = H(\delta, - \vec k)$ and the Bloch wavefunction projected to each spin state transforms as 
\be 
\epsilon^s_n (\delta, \vec k) = \epsilon^{-s}_n (\delta, - \vec k), \quad \mathcal T | u^{s}_n (\delta, \vec k) \ra = | u^{-s}_n (\delta, - \vec k) \ra^*, 
\ee
where $s = \uparrow, \downarrow$ denotes the $z$ component of the electron spin (arising from the large Ising spin-orbit coupling in the valleys of TMDs). The above wavefunction transformation yields
\be
\la u^{s_1}_m (\delta, \vec k_1) | u^{s_2}_n (\delta, \vec k_2) \ra = \la u^{s_1}_m (\delta, \vec k_1) | \mathcal T^{-1} \mathcal T | u^{s_2}_n (\delta, \vec k_2) \ra = \la u^{-s_1}_m (\delta,  - \vec k_1) | u^{-s_2}_n (\delta,  -\vec k_2) \ra^*. 
\ee
The velocity operator transforms as $\mathcal T \hat \nu(\delta) \mathcal T^{-1} = - \hat \nu (\delta)$. Thus, for a given polarisation angle $\theta$, the velocity matrix element satisfies 
\begin{align} \label{Seq:TMD_nu}
\la u^{s_1}_m (\delta, \vec k_1) | \nu_\theta (\delta) | u^{s_2}_n (\delta, \vec k_2) \ra &= \la u^{s_1}_m (\delta, \vec k_1) | \mathcal T^{-1} \mathcal T \nu_\theta (\delta) \mathcal T^{-1} \mathcal T | u^{s_2}_n (\delta, \vec k_2) \ra \nonumber \\
&= - \la u^{-s_1}_m (\delta, -\vec k_1) | \nu_\theta (\delta) | u^{-s_2}_n (\delta, -\vec k_2) \ra^*. 
\end{align}
We obtain the symmetry constraint for the Wilson line: 
\be 
\mathcal W^s (\delta, \theta, \vec k, \vec q) = - [\mathcal W^{-s} (\delta, \theta, -\vec k, - \vec q)]^*, \quad \arg [ \mathcal W^s (\delta, \theta, \vec k, \vec q)] = - \arg [ \mathcal W^{-s} (\delta, \theta, - \vec k, -\vec q)] + \pi.
\ee
As a result, the spin-dependent shift vector in a time-reversal invariant system satisfies 
\be\label{Seq:TMD_T} 
\vec r^s (\delta, \theta, \vec k) =  \vec r^{-s} (\delta, \theta, -\vec k). 
\ee
Due to the large (Ising) spin-valley locking in TMDs for photon energies close to the bandgap, we anticipate that this spin SPC is locked to each of the valleys. However, the spin-resolved shift vectors have the same sign [see Eq.~(\ref{Seq:TMD_T}), albeit at opposite $\vec k$]. Since the square of the velocity matrix element is even under $\vec k \to - \vec k$ and $s \to -s$ [Eq.~(\ref{Seq:TMD_nu})], the weighted shift vector satisfies $\vec R^s (\delta, \theta, \vec k) = \vec R^{-s} (\delta, \theta, -\vec k)$.
Thus, upon integration over the $k$-space, both $\uparrow,\downarrow$ spin SPC move in the same direction yielding a charge current.

\subsubsection{Inversion}
Due to the sublattice potential difference $\delta$, monolayer TMDs also break inversion symmetry. Under spatial inversion operation, we have $\mathcal I \mathcal H (\delta, \vec r) \mathcal I^{-1} =  \mathcal H (-\delta, \vec r) $ and $\mathcal I H (\delta, \vec k) \mathcal I^{-1} = H (-\delta, -\vec k)$. On the other hand, spin is invariant under inversion. Thus the spin resolved Bloch wavefunction satisfies
\be
\epsilon^s_n (\delta, \vec k) = \epsilon^s_n (-\delta, - \vec k), \quad \mathcal I |u^s_n (\delta, \vec k) \ra = | u^s_n (-\delta, -\vec k) \ra.  
\ee
Thus we have 
\be
\la u^{s_1}_m (\delta, \vec k_1) | u^{s_2}_n (\delta, \vec k_2) \ra = \la u^{s_1}_m (\delta, \vec k_1) | \mathcal I^{-1} \mathcal I | u^{s_2}_n (\delta, \vec k_2) \ra = \la u^{s_1}_m (-\delta,  - \vec k_1) | u^{s_2}_n (-\delta,  -\vec k_2) \ra. 
\ee

The velocity operator transforms as $\mathcal I \hat \nu(\delta) \mathcal I^{-1} = - \hat \nu (-\delta)$. This yields
\begin{align}
\la u^{s_1}_m (\delta, \vec k_1) | \nu_\theta (\delta) | u^{s_2}_n (\delta, \vec k_2) \ra &= \la u^{s_1}_m (\delta, \vec k_1) | \mathcal I^{-1} \mathcal I \nu_\theta (\delta) \mathcal I^{-1} \mathcal I | u^{s_2}_n (\delta, \vec k_2) \ra \nonumber \\
&= - \la u^{s_1}_m (-\delta, -\vec k_1) | \nu_\theta (-\delta) | u^{s_2}_n (-\delta, -\vec k_2) \ra. 
\end{align}
We obtain the symmetry constraint for the Wilson line: 
\be 
\mathcal W^s (\delta, \theta, \vec k, \vec q) = - \mathcal W^s (-\delta, \theta, -\vec k, - \vec q), \quad \arg [ \mathcal W^s (\delta, \theta, \vec k, \vec q)] =  \arg [ \mathcal W^s (-\delta, \theta, - \vec k, -\vec q)] + \pi. 
\ee
As a result, the shift vector obeys the following relation:
\be\label{Seq:TMD_I} 
\vec r^s (\delta, \theta, \vec k) = - \vec r^s (-\delta, \theta, -\vec k). 
\ee
The spin SPC can be obtained by integrating the weighted shift vector over the entire $k$-space [Eq.~(2) in the main text], thus Eq.~(\ref{Seq:TMD_I}) requires that the spin resolved SPC reverses sign for $\delta \to - \delta$. Furthermore, the charge current can be calculated by summing over the spin SPC and also flips sign for $\delta \to - \delta$.

\subsubsection{Mirror Symmetry}
Monolayer TMD also exhibits mirror symmetry about the $y$-axis: $\mathcal M_y \mathcal H (\delta, \vec r) \mathcal M_y^{-1} = \mathcal H(\delta, \vec r)$ and $\mathcal M_y H (\delta, \vec k) \mathcal M_y^{-1} = H (\delta, \mathcal M_y \vec k)$. On the other hand, spin transforms in the same way as angular momentum upon reflection. In 3D, for a $yz$ mirror plane, the $z$ component of the spin flips sign under $\mathcal M_y: (x,y,z) \to (-x, y, -z)$. Thus the wavefunction projected to each spin satisfies  
\be
\epsilon^s_n (\delta, \vec k) = \epsilon^{-s}_n (\delta, \mathcal M_y \vec k), \quad \mathcal M_y |u^s_n (\delta, \vec k) \ra = | u^{-s}_n (\delta, \mathcal M_y \vec k) \ra,  
\ee
and
\be
\la u^{s_1}_m (\delta, \vec k_1) | u^{s_2}_n (\delta, \vec k_2) \ra = \la u^{s_1}_m (\delta, \vec k_1) | \mathcal M_y^{-1} \mathcal M_y | u^{s_2}_n (\delta, \vec k_2) \ra = \la u^{-s_1}_m (\delta,  \mathcal M_y \vec k_1) | u^{-s_2}_n (\delta,  \mathcal M_y \vec k_2) \ra. 
\ee

The velocity operators transform as $\mathcal M_y \nu_x (\delta)  \mathcal M_y^{-1} = - \nu_x (\delta) $ and $\mathcal M_y \nu_y (\delta)  \mathcal M_y^{-1} = \nu_y (\delta) $, thus we have $\mathcal M_y \nu_\theta (\delta)  \mathcal M_y^{-1} =  \nu_{\pi - \theta} (\delta) $. This gives 
\begin{align}
\la u^{s_1}_m  (\delta, \vec k_1) | \nu_\theta (\delta)  | u^{s_2}_n (\delta, \vec k_2) \ra &= \la u^{s_1}_m  (\delta, \vec k_1) | \mathcal M_y^{-1} \mathcal M_y \nu_\theta (\delta)  \mathcal M_y^{-1} \mathcal M_y | u^{s_2}_n  (\delta, \vec k_2) \ra \nonumber \\
&= \la u^{-s_1}_m (\delta, \mathcal M_y \vec k_1) | \nu_{\pi - \theta} (\delta)  | u^{-s_2}_n (\delta, \mathcal M_y \vec k_2) \ra. 
\end{align}
The spin-dependent Wilson line satisfies 
\be
\mathcal W^s (\delta, \theta, \vec k, \vec q) = \mathcal W^{-s} (\delta, \pi - \theta, \mathcal M_y  \vec k, \mathcal M_y \vec q), \quad \arg [ \mathcal W^s (\delta, \theta, \vec k,  \vec q)] = \arg [ \mathcal W^{-s} (\delta, \pi- \theta, \mathcal M_y \vec k, \mathcal M_y \vec q)]  , 
\ee
and shift vector obeys
\be\label{Seq:TMD_My_intermediate}
r^s_x  (\delta, \theta, \vec k) = - r^{-s}_x  (\delta, \pi - \theta, \mathcal M_y \vec k) , \quad r^s_y  (\delta, \theta, \vec k) = r^{-s}_y  (\delta, \pi - \theta, \mathcal M_y \vec k). 
\ee
Since the polarisation directions $\theta$ and $\theta +\pi$ are equivalent, Eq.~(\ref{Seq:TMD_My_intermediate}) as 
\be\label{Seq:TMD_My}
r^s_x  (\delta, \theta, \vec k) = - r^{-s}_x  (\delta,  - \theta, \mathcal M_y \vec k) , \quad r^s_y  (\delta, \theta, \vec k) = r^{-s}_y  (\delta, - \theta, \mathcal M_y \vec k). 
\ee
Combining with Eq.~(\ref{Seq:TMD_T}), we obtain the shift vector symmetry constraint for each spin: 
\be\label{Seq:TMD_TMy}
r^s_x  (\delta, \theta, \vec k) = - r^{s}_x  (\delta,  - \theta, k_x, -k_y) , \quad r^s_y  (\delta, \theta, \vec k) = r^{s}_y  (\delta, - \theta, k_x, -k_y). 
\ee

Similar to the charge current in BLG and G/hBN, the spin SPC is sensitive to the linear polarisation of light. For electric field polarised perpendicular to ($\theta = 0,\pi$) or parallel with ($\theta = \pm \pi/2$) the mirror axis, $r_x^s (\delta, \theta, \vec k)$ flips sign for $k_y \to - k_y$ while $r_y^s (\delta, \theta, \vec k)$ is invariant. This leads to transverse (longitudinal) spin SPC for polarisation perpendicular to (parallel with) the mirror axis. Since Eq.~(\ref{Seq:TMD_TMy}) is valid for both spins, we anticipate that the charge SPC also exhibits the same polarisation dependence, i.e. the charge SPC is transverse (longitudinal) for electric field polarisation normal to (parallel with) the mirror plane. 

Furthermore, monolayer TMDs are also invariant under $C^z_3$ and possesses three mirror reflection axes that are separated by $\pm 2 \pi/3$ apart from each other. Thus a similar argument on the polarisation dependent spin and charge SPC can be extended to other mirror reflection axes after rotating by $\pm 2 \pi /3$ about the $z$-axis.

\subsection{Shift Vector Configuration Dependence for 2H Stacked Bilayer TMDs}
SPC in bilayer TMDs exhibits stacking and polarisation dependence as well. Here we focus on the most common 2H stacking configuration for bilayer TMDs, whereby one of the layers is rotated by $\pi$ about the $z$-axis and then directly stacked on top of the other. In this stacking configuration, the A (B) site of the top layer is directly on top of the B (A) site of the bottom layer. We describe the system with a real space Hamiltonian $\mathcal H(\Delta, \delta, \vec r)$, where $\Delta$ is the interlayer potential difference provided by an external out-of-plane electric field, and $\delta$ is the sublattice potential difference between A and B sites. In the absence of interlayer potential difference, the material is centrosymmetric and the SPC vanishes. Here we consider the symmetry constraints of the spin dependent shift vector and SPC when $\Delta \neq 0$.

\subsubsection{Time Reversal Symmetry}
Time reversal symmetry demands $\mathcal T H (\Delta, \delta, \vec k) \mathcal T^{-1} = H(\Delta, \delta, - \vec k)$. Thus the wavefunction projected to each spin satisfies 
\be 
\epsilon^s_n (\Delta, \delta, \vec k) = \epsilon^{-s}_n (\Delta, \delta, - \vec k), \quad \mathcal T | u^{s}_n (\Delta, \delta, \vec k) \ra = | u^{-s}_n (\Delta, \delta, - \vec k) \ra^*, 
\ee
and 
\be
\la u^{s_1}_m (\Delta, \delta, \vec k_1) | u^{s_2}_n (\Delta, \delta, \vec k_2) \ra = \la u^{s_1}_m (\Delta, \delta, \vec k_1) | \mathcal T^{-1} \mathcal T | u^{s_2}_n (\Delta, \delta, \vec k_2) \ra = \la u^{-s_1}_m (\Delta, \delta,  - \vec k_1) | u^{-s_2}_n (\Delta, \delta,  -\vec k_2) \ra^*. 
\ee
Here, $s = \uparrow, \downarrow$ denotes the the electron spin along the $z$-axis. 

The velocity operator transforms as $\mathcal T \hat \nu(\Delta, \delta) \mathcal T^{-1} = - \hat \nu (\Delta, \delta)$. Thus, for a given polarisation angle $\theta$, the velocity matrix element satisfies 
\begin{align}
\la u^{s_1}_m (\Delta, \delta, \vec k_1) | \nu_\theta (\Delta, \delta) | u^{s_2}_n (\Delta, \delta, \vec k_2) \ra &= \la u^{s_1}_m (\Delta, \delta, \vec k_1) | \mathcal T^{-1} \mathcal T \nu_\theta (\Delta, \delta) \mathcal T^{-1} \mathcal T | u^{s_2}_n (\Delta, \delta, \vec k_2) \ra \nonumber \\
&= - \la u^{-s_1}_m (\Delta, \delta, -\vec k_1) | \nu_\theta (\Delta, \delta) | u^{-s_2}_n (\Delta, \delta, -\vec k_2) \ra^*. 
\end{align}
We obtain the symmetry constraint for the Wilson line: 
\be 
\mathcal W^s (\Delta, \delta, \theta, \vec k, \vec q) = - [\mathcal W^{-s} (\Delta, \delta, \theta, -\vec k, - \vec q)]^*, \quad \arg [ \mathcal W^s (\Delta, \delta, \theta, \vec k, \vec q)] = - \arg [ \mathcal W^{-s} (\Delta, \delta, \theta, - \vec k, -\vec q)] + \pi.
\ee
As a result, the spin-dependent shift vector satisfies 
\be\label{Seq:2H_T} 
\vec r^s (\Delta, \delta, \theta, \vec k) =  \vec r^{-s} (\Delta, \delta, \theta, -\vec k). 
\ee
 Similar with the scenario in a monolayer TMD as discussed in the previous subsection, Eq.~(\ref{Seq:2H_T}) leads to spin SPC that is locked to each valley. We also note that the (weighted) shift vectors for opposite spins have the same sign at opposite $\vec k$. Thus the spin currents for $s = \uparrow, \downarrow$ flow in the same direction, leading to a net charge SPC.

\subsubsection{Inversion}
Under spatial inversion, the atomic configuration of the 2H stacked bilayer TMD remains invariant, while the interlayer potential flips sign: $\mathcal I \mathcal H (\Delta, \delta, \vec r) \mathcal I^{-1} = \mathcal H (- \Delta, \delta, \vec r)$. The Bloch Hamiltonian thus satisfies $\mathcal I H (\Delta, \delta, \vec k) \mathcal I^{-1} = H (- \Delta, \delta, -\vec k)$. This gives 
\be 
\epsilon^s_n (\Delta, \delta, \vec k) = \epsilon^{s}_n (-\Delta, \delta, - \vec k), \quad \mathcal I | u^{s}_n (\Delta, \delta, \vec k) \ra = | u^{s}_n (-\Delta, \delta, - \vec k) \ra, 
\ee
and 
\be
\la u^{s_1}_m (\Delta, \delta, \vec k_1) | u^{s_2}_n (\Delta, \delta, \vec k_2) \ra = \la u^{s_1}_m (\Delta, \delta, \vec k_1) | \mathcal I^{-1} \mathcal I | u^{s_2}_n (\Delta, \delta, \vec k_2) \ra = \la u^{s_1}_m (- \Delta, \delta,  - \vec k_1) | u^{s_2}_n (- \Delta, \delta,  -\vec k_2) \ra.
\ee

Upon spatial inversion, the velocity operator transforms as $\mathcal I \hat \nu(\Delta, \delta) \mathcal I^{-1} = - \hat \nu (-\Delta, \delta)$. This gives 
\begin{align}
\la u^{s_1}_m (\Delta, \delta, \vec k_1) | \nu_\theta (\Delta, \delta) | u^{s_2}_n (\Delta, \delta, \vec k_2) \ra &= \la u^{s_1}_m (\Delta, \delta, \vec k_1) | \mathcal I^{-1} \mathcal I \nu_\theta (\Delta, \delta) \mathcal I^{-1} \mathcal I | u^{s_2}_n (\Delta, \delta, \vec k_2) \ra \nonumber \\
&= - \la u^{s_1}_m (-\Delta, \delta, -\vec k_1) | \nu_\theta (-\Delta, \delta) | u^{s_2}_n (-\Delta, \delta, -\vec k_2) \ra. 
\end{align}
It follows that the Wilson line satisfies
\be 
\mathcal W^s (\Delta, \delta, \theta, \vec k, \vec q) = - \mathcal W^{s} (-\Delta, \delta, \theta, -\vec k, - \vec q), \quad \arg [ \mathcal W^s (\Delta, \delta, \theta, \vec k, \vec q)] = \arg [ \mathcal W^{s} (-\Delta, \delta, \theta, - \vec k, -\vec q)] + \pi.
\ee
As a result, the spin-dependent shift vector satisfies 
\be\label{Seq:2H_I} 
\vec r^s (\Delta, \delta, \theta, \vec k) =  -\vec r^{s} (-\Delta, \delta, \theta, -\vec k). 
\ee

Since the SPC is obtained by integrating the weighted shift vector and both $\rho(\Delta, \delta, \vec k)$ and $| \nu_\theta (\Delta, \delta, \vec k)|^2$ are even in $\Delta$ and $\vec k$, Eq.~(\ref{Seq:2H_I}) implies that the spin SPC flips direction for $\Delta \to - \Delta$. The charge SPC can be obtained by summing over $s= \uparrow, \downarrow$ and thus also reverses its direction upon reversing interlayer potential difference.

\subsubsection{Mirror Symmetry}

2H bilayer TMD is invariant under mirror reflection about the $y$ axis: $\mathcal M_y H (\Delta, \delta, \vec k) \mathcal M_y^{-1} = H (\Delta, \delta, \mathcal M_y \vec k)$. The spin acts as a pseudovector upon reflection, and thus $s$ flips sign under $\mathcal M_y$. This leads to 
\be
\epsilon^s_n (\Delta, \delta, \vec k) = \epsilon^{-s}_n (\Delta, \delta, \mathcal M_y \vec k), \quad \mathcal M_y | u^{s}_n (\Delta, \delta, \vec k) \ra = | u^{-s}_n (\Delta, \delta, \mathcal M_y \vec k) \ra, 
\ee
and
\be
\la u^{s_1}_m (\Delta, \delta, \vec k_1) | u^{s_2}_n (\Delta, \delta, \vec k_2) \ra = \la u^{s_1}_m (\Delta, \delta, \vec k_1) | \mathcal M_y^{-1} \mathcal M_y | u^{s_2}_n (\Delta, \delta, \vec k_2) \ra = \la u^{-s_1}_m (\Delta, \delta,  \mathcal M_y \vec k_1) | u^{-s_2}_n (\Delta, \delta,  \mathcal M_y \vec k_2) \ra.
\ee

The velocity operators transform as $\mathcal M_y \nu_x (\Delta, \delta)  \mathcal M_y^{-1} = - \nu_x (\Delta, \delta) $ and $\mathcal M_y \nu_y (\Delta, \delta)  \mathcal M_y^{-1} = \nu_y (\Delta, \delta) $, thus we have $\mathcal M_y \nu_\theta (\Delta, \delta) \mathcal M_y^{-1} =  \nu_{\pi - \theta} (\Delta, \delta) $. This gives 
\begin{align}
\la u^{s_1}_m  (\Delta, \delta, \vec k_1) | \nu_\theta (\Delta, \delta)  | u^{s_2}_n (\Delta, \delta, \vec k_2) \ra &= \la u^{s_1}_m  (\Delta, \delta, \vec k_1) | \mathcal M_y^{-1} \mathcal M_y \nu_\theta (\Delta, \delta)  \mathcal M_y^{-1} \mathcal M_y | u^{s_2}_n  (\Delta, \delta, \vec k_2) \ra \nonumber \\
&= \la u^{-s_1}_m (\Delta, \delta, \mathcal M_y \vec k_1) | \nu_{\pi - \theta} (\Delta, \delta)  | u^{-s_2}_n (\Delta, \delta, \mathcal M_y \vec k_2) \ra. 
\end{align}
The spin-dependent Wilson line satisfies 
\be
\mathcal W^s (\Delta, \delta, \theta, \vec k, \vec q) = \mathcal W^{-s} (\Delta, \delta, \pi - \theta, \mathcal M_y  \vec k, \mathcal M_y \vec q), \quad \arg [ \mathcal W^s (\Delta, \delta, \theta, \vec k,  \vec q)] = \arg [ \mathcal W^{-s} (\Delta, \delta, \pi- \theta, \mathcal M_y \vec k, \mathcal M_y \vec q)]  , 
\ee
and shift vector obeys
\be\label{Seq:2H_My_intermediate}
r^s_x  (\Delta, \delta, \theta, \vec k) = - r^{-s}_x  (\Delta, \delta, \pi - \theta, \mathcal M_y \vec k) , \quad r^s_y  (\Delta, \delta, \theta, \vec k) = r^{-s}_y  (\Delta, \delta, \pi - \theta, \mathcal M_y \vec k). 
\ee
Again, we use the identity that the polarisation directions $\theta$ and $\theta +\pi$ are equivalent to obtain 
\be\label{Seq:2H_My}
r^s_x (\Delta, \delta, \theta, \vec k) = - r^{-s}_x  (\Delta, \delta,  - \theta, \mathcal M_y \vec k) , \quad r^s_y  (\Delta, \delta, \theta, \vec k) = r^{-s}_y  (\Delta, \delta, - \theta, \mathcal M_y \vec k). 
\ee
Combining with Eq.~(\ref{Seq:2H_T}), we obtain the shift vector symmetry constraint for each spin: 
\be\label{Seq:2H_TMy}
r^s_x  (\Delta, \delta, \theta, \vec k) = - r^{s}_x  (\Delta, \delta,  - \theta, k_x, -k_y) , \quad r^s_y  (\Delta, \delta, \theta, \vec k) = r^{s}_y  (\Delta, \delta, - \theta, k_x, -k_y). 
\ee

Eq.~(\ref{Seq:2H_TMy}) ensures that for polarisation angles $\theta = 0, \pi$ and $\theta = \pm \pi/2$, $r^s_x  (\Delta, \delta, \theta, \vec k)$ is odd under $k_y \to - k_y$ while $r^s_y  (\Delta, \delta, \theta, \vec k)$ is even. Thus, for electric field polarised perpendicular to (parallel with) with mirror reflection axis, the spin SPC is transverse (longitudinal). Again, similar with the scenario in a monolayer TMD, since the symmetry constraint in Eq.~(\ref{Seq:2H_TMy}) is valid for both spins, the charge SPC exhibits the same polarisation dependence as the spin SPC. 

We remark that 2H stacked bilayer TMD also possesses three-fold in-plane rotational symmetry $C^z_3$ and thus has two more mirror axes that are separated $\pm 2\pi/3$ from the $y$-axis. The argument above on polarisation dependent spin and charge SPC can also be applied to the other two mirror axes after rotating the coordinate system by $\pm 2 \pi/3$.

Now we move on to $\mathcal M_x$. 2H stacked bilayer TMD breaks mirror symmetry with respect to the $x$-axis. $\mathcal M_x$ interchanges the A, B sublattice sites, thus reversing $\delta$: $\mathcal M_x \mathcal H (\Delta, \delta, \vec r) \mathcal M_x = \mathcal H (\Delta, -\delta, \vec r)$. The Bloch Hamiltonian satisfies $\mathcal M_x H (\Delta, \delta, \vec k) \mathcal M_x = H (\Delta, -\delta, \mathcal M_x \vec k)$. For wavefunctions projected to each spin state, under mirror reflection, the spin transforms as a pseudovector and the $z$ component $s$ flips sign. Thus we have 
\be
\epsilon^s_n (\Delta, \delta, \vec k) = \epsilon^{-s}_n (\Delta, -\delta, \mathcal M_x \vec k), \quad \mathcal M_x | u^{s}_n (\Delta, \delta, \vec k) \ra = | u^{-s}_n (\Delta, -\delta, \mathcal M_x \vec k) \ra, 
\ee
and
\be
\la u^{s_1}_m (\Delta, \delta, \vec k_1) | u^{s_2}_n (\Delta, \delta, \vec k_2) \ra = \la u^{s_1}_m (\Delta, \delta, \vec k_1) | \mathcal M_x^{-1} \mathcal M_x | u^{s_2}_n (\Delta, \delta, \vec k_2) \ra = \la u^{-s_1}_m (\Delta, -\delta,  \mathcal M_x \vec k_1) | u^{-s_2}_n (\Delta, -\delta,  \mathcal M_x \vec k_2) \ra.
\ee 

Under $\mathcal M_x$, the $y$ component of the velocity operator switches sign $\mathcal M_x \nu_y (\Delta, \delta)  \mathcal M_x^{-1} = - \nu_y (\Delta, -\delta) $ while the $x$ component is invariant $\mathcal M_x \nu_x (\Delta, \delta)  \mathcal M_x^{-1} = \nu_x (\Delta, -\delta) $. Thus for a given polarisation $\theta$, we have $\mathcal M_x \nu_\theta (\Delta, \delta)  \mathcal M_x^{-1} =  \nu_{-\theta} (\Delta, -\delta) $ and 
\begin{align}
\la u^{s_1}_m  (\Delta, \delta, \vec k_1) | \nu_\theta (\Delta, \delta)  | u^{s_2}_n (\Delta, \delta, \vec k_2) \ra &= \la u^{s_1}_m  (\Delta, \delta, \vec k_1) | \mathcal M_x^{-1} \mathcal M_x \nu_\theta (\Delta, \delta)  \mathcal M_x^{-1} \mathcal M_x | u^{s_2}_n  (\Delta, \delta, \vec k_2) \ra \nonumber \\
&= \la u^{-s_1}_m (\Delta, -\delta, \mathcal M_x \vec k_1) | \nu_{- \theta} (\Delta, -\delta)  | u^{-s_2}_n (\Delta, -\delta, \mathcal M_x \vec k_2) \ra. 
\end{align}
Thus the Wilson line satisfies 
\be
\mathcal W^s (\Delta, \delta, \theta, \vec k, \vec q) = \mathcal W^{-s} (\Delta, -\delta, - \theta, \mathcal M_x  \vec k, \mathcal M_x \vec q), \quad \arg [ \mathcal W^s (\Delta, \delta, \theta, \vec k,  \vec q)] = \arg [ \mathcal W^{-s} (\Delta, -\delta, - \theta, \mathcal M_x \vec k, \mathcal M_x \vec q)]  , 
\ee
The spin dependent shift vector satisfies 
\be\label{Seq:2H_Mx}
r^s_x  (\Delta, \delta, \theta, \vec k) = r^{-s}_x  (\Delta, -\delta, - \theta, \mathcal M_x \vec k) , \quad r^s_y  (\Delta, \delta, \theta, \vec k) = - r^{-s}_y  (\Delta, -\delta, - \theta, \mathcal M_x \vec k). 
\ee

Combining Eq.~(\ref{Seq:2H_My}) and~(\ref{Seq:2H_Mx}), we obtain the dependence of the spin shift vector on $\delta$: 
\be\label{Seq:2H_delta}
\vec r^s (\Delta, \delta, \theta, \vec k) =  -\vec r^{s} (\Delta, -\delta, \theta, -\vec k). 
\ee
We observe that the sublattice potential difference $\delta$ plays a similar role with the interlayer potential difference $\Delta$ in Eq.~(\ref{Seq:2H_I}).

\section{Strained Shift Photocurrent Induced by Unpolarised Light}

In this section, we show that in strained vdW materials and homostructures, 
the SPC induced by unpolarised light is nonzero due to broken discrete rotational symmetry. For brevity, we will omit the implicit dependence of the shift vector $\vec r (\theta, \vec k)$ in stacking configuration index $\eta$, spin $s$, interlayer potential difference $\Delta$ and sublattice potential difference $\delta$ in this section. 

The strained SPC induced by unpolarised light can be calculated by integrating $\vec j (\theta)$ all distinct polarisation angles: 
\be\label{Seq:unpolarised1}
\vec J^{\rm tot} = \int_{-\pi/2}^{\pi/2} \vec j (\theta) d \theta = C \int_{- \pi/2}^{\pi/2} d \theta \int d \vec k  \rho (\vec k) |\nu_\theta (\vec k) |^2 \vec r (\theta, \vec k).
\ee
We can write $|\nu_\theta (\vec k)|^2 = V_s (\theta, \vec k) + V_a (\theta, \vec k)$, where $V_s (\theta, \vec k) = |\nu_x (\vec k)|^2 \cos^2 \theta + |\nu_y (\vec k)|^2 \sin^2 \theta$ is the component symmetric with respect to $\theta \to - \theta$, and $V_a (\theta, \vec k) = 2\Re \left[ \nu_x (\vec k) \nu_y^* (\vec k)\right] \sin \theta \cos \theta$ is the component antisymmetric with respect to $\theta \to - \theta$. Also, we note that $\rho(\vec k)$ is only dependent on the dispersion and is thus symmetric in $\theta$. 
Then the integration in Eq.~(\ref{Seq:unpolarised1}) can be rewritten as
\be\label{Seq:unpolarised2}
\vec J^{\rm tot} =C \int_0^{\pi/2} d \theta \int d \vec k \rho(\vec k) \left[ V_s (\theta, \vec k) \left( \vec r (\theta, \vec k) + \vec r (-\theta, \vec k)\right) + V_a (\theta, \vec k) \left( \vec r (\theta, \vec k) - \vec r (- \theta, \vec k)\right) \right].
\ee 

When strain is applied either parallel or perpendicular to a mirror axis, mirror symmetry about the mirror axis is preserved. For instance, when strain is applied perpendicular or parallel to $y$ in any of the vdW systems discussed above (Bernal stacked BLG, G/hBN, monolayer TMD and 2H stacked bilayer TMD), time reversal symmetry $\mathcal T$ and mirror symmetry $\mathcal M_y$ are preserved. As we now show, this symmetry dramatically constrain the integrand of Eq.~(\ref{Seq:unpolarised2}).

To see this, we examine $J^{\rm tot}_x$ and $J^{\rm tot}_y$ separately. We first concentrate on $J^{\rm tot}_x$. We note that mirror symmetry $y$ [in particular, the first equation in Eq.~(\ref{Seq:TMy})] guarantees that $r_x (\theta, \vec k) + r_x (-\theta, \vec k)$ is odd in $k_y$ while $V_s (\theta, \vec k)$ is even in $k$-space. Thus the term proportional to $V_s(\theta, \vec k)$ in the integration in Eq.~(\ref{Seq:unpolarised2}) vanishes. Furthermore, $r_x (\theta, \vec k) -  r_x (- \theta, \vec k)$ is even in $k_y$ while $V_a (\theta, \vec k)$ is odd [since $\nu_y (\vec k)$ is odd while $\nu_x(\vec k)$ is even]. Thus the term proportional to $V_a (\theta, \vec k)$ also vanishes after integration. Therefore, $J^{\rm tot}_x = 0$ in the presence of $\mathcal T$ and $\mathcal M_y$. 

We now employ a similar argument for $J^{\rm tot}_y$. We note that that mirror symmetry about $y$ [in particular, the second equation in Eq.~(\ref{Seq:TMy})] guarantees that $r_y (\theta, \vec k) + r_y (-\theta, \vec k)$ is even in $k_y$ so that the term proportional to $V_s (\theta, \vec k)$ is even. Furthermore, $r_y (\theta, \vec k) -  r_y (- \theta, \vec k)$ is odd in $k_y$ so that the term proportional to $V_a (\theta, \vec k)$ is also even in $k_y$. Thus $J^{\rm tot}_y$ is finite and the SPC induced by unpolarised light is directed along the $y$ axis. As a result, when strain is applied either perpendicular or parallel to a mirror axis, $\vec J^{\rm tot}$ is directed along the mirror axis, as discussed in the main text. However, when strain is not applied either perpendicular or parallel to a mirror axis, $\vec J^{\rm tot}$ does not generically point in a symmetry determined direction; instead its direction is determined by the details of the strain, as discussed in the main text.

Importantly, this argument (presented above) for the integrated SPC over all polarizations (i.e. SPC for unpolarized light) applies for in the presence of any (and multiple) mirror planes. For instance, in the absence of strain, the vdW materials and homostructures discussed above possess an in-plane rotational symmetry $C^z_3$. This means that vdW materials discussed above possess multiple non-parallel mirror planes. Upon irradiation of unpolarised light and in an unstrained system, it is impossible for $\vec J^{\rm tot}$ to be directed along all these non-parallel reflection axes simultaneously, and thus the integrated SPC has to vanish.

\section{Hamiltonian of Bernal Stacked BLG}

In this section, we derive the low-energy four-band Hamiltonian for Bernal stacked BLG from the tight-binding model. Bernal stacked BLG (Fig.~1c in the main text) has a triangular lattice with primitive lattice vectors given by 
\be
\vec a_1 = \left( \frac{a}{2}, \; \frac{\sqrt{3}a}{2} \right), \quad \vec a_2 =  \left( \frac{a}{2}, \; -\frac{\sqrt{3}a}{2} \right), 
\ee
where $a = 0.246 \; {\rm nm}$. 
Each unit cell contains four atoms, two from the top layer ($\rm A_t$, $\rm B_t$) and two from the bottom layer ($\rm A_b$ and $\rm B_b$). In AB stacking configuration, the sublattice site of the top layer $\rm A_t$ is directly on top of the sublattice site $\rm B_b$ of the bottom layer (referred to as the dimer site), while site $\rm B_t$ sits at the centre of the honeycomb lattice of the bottom layer (referred to as the non-dimer site). In BA stacking. we have $\rm B_t$ directly on top of $\rm A_b$ instead. As we remarked in the previous section, these two stacking configurations are obtained by flipping one about the in-plane axis and they have the same dispersion relations. 

Now we derive the four band tight-binding Hamiltonian of the system $H^{(\eta)}$ in the basis $\{ \psi_{\rm A_b}, \psi_{\rm B_b}, \psi_{\rm A_t}, \psi_{\rm B_t} \}$. When an interlayer electric potential difference is applied, the top and bottom lattice sites acquire different onsite energies leading to nonzero on-diagonal terms in the tight-bindind Hamiltonian: $H^{(\eta)}_{\rm A_b \rm A_b} = H^{(\eta)}_{\rm B_b \rm B_b} = \Delta/2$ and $H^{(\eta)}_{\rm A_t \rm A_t} = H^{(\eta)}_{\rm B_t \rm B_t} = -\Delta/2$. 

Furthermore, the intralayer nearest-neighbour hopping from site A to site B is described the vectors 
\be
\boldsymbol{\delta}_1 = \left( 0, \; \frac{a}{\sqrt{3}} \right), \quad \boldsymbol{\delta}_2= \left( \frac{a}{2}, \; -\frac{a}{2\sqrt{3}} \right), \quad \boldsymbol{\delta}_3 = \left( -\frac{a}{2}, \; -\frac{a}{2\sqrt{3}} \right). 
\ee
Thus, the intralayer hopping in each layer is described by $H^{(\eta)}_{\rm A_b \rm B_b} = H^{(\eta)}_{\rm A_t \rm B_t} = - \gamma_0 f(\vec k)$, where $\gamma_0 \approx 3 \; {\rm eV}$ is the nearest neighbour hopping constant and 
\begin{align}
f(\vec k) &= \sum_{l=1}^{3} e^{i \vec k \cdot \boldsymbol{\delta}_l} = e^{i \frac{a}{\sqrt{3}} k_y } + e^{i(\frac{a}{2} k_x - \frac{a}{2 \sqrt{3}} k_y)} + e^{i(-\frac{a}{2} k_x - \frac{a}{2 \sqrt{3}} k_y)} \nonumber \\
&= e^{i \frac{a}{\sqrt{3}} k_y } + 2 e^{- i \frac{a}{2 \sqrt{3}} k_y} \cos{\left( \frac{a}{2} k_x \right)}.
\end{align}

In AB stacking, since ${\rm A_t}$ is directly on top of $\rm B_b$, the interlayer coupling at the dimer site is given by $H^{(\rm AB)}_{\rm A_t B_b} = H^{(\rm AB)}_{\rm B_b A_t} = \gamma_1$. The skew interlayer hopping from $\rm A_b$ to $\rm B_t$ at the non-dimer sites involves in-plane hopping described by the in-plane vectors $-\boldsymbol{\delta}_1$, $ -\boldsymbol{\delta}_2$ and $ -\boldsymbol{\delta}_3$, thus the skew interlayer coupling is given by $H^{(\rm AB)}_{\rm A_b B_t} = - \gamma_3 \sum_{l=1}^{3} e^{i \vec k \cdot (-\boldsymbol{\delta}_l)} = -\gamma_3 f^* (\vec k)$. The next next nearest neighbour interlayer hopping between $\rm A_b$ and $\rm A_t$ as well as between $\rm B_b$ and $\rm B_t$ is much smaller than the other terms and can be neglected here. Therefore, we arrive at the tight-binding Hamiltonian for the AB stacked BLG:
\be 
H^{(\rm AB)} (\vec k) = \begin{pmatrix}
\frac{\Delta}{2} & -\gamma_0 f(\vec k) & 0 & -\gamma_3 f^* (\vec k) \\ 
-\gamma_0 f^* (\vec k) & \frac{\Delta}{2} & \gamma_1  & 0 \\
0 & \gamma_1 & -\frac{\Delta}{2} & -\gamma_0 f(\vec k) \\
-\gamma_3 f(\vec k) & 0 & -\gamma_0 f^* (\vec k) & -\frac{\Delta}{2}
\end{pmatrix}
\ee

In BA stacking, $\rm B_t$ is directly on top of $\rm A_b$, and the interlayer coupling at the dimer site is $H^{(\rm BA)}_{\rm B_t A_b} = H^{(\rm BA)}_{\rm A_b B_t} = \gamma_1$. The skew interlayer hopping from $\rm B_b$ to $\rm A_t$ is given by the in-plane vectors $\boldsymbol{\delta}_1$, $\boldsymbol{\delta}_2$ and $\boldsymbol{\delta}_3$, yielding $H^{(\rm BA)}_{\rm B_b A_t} = -\gamma_3 f (\vec k)$. Thus the Hamiltonian for the BA stacked BLG is 
\be
H^{(\rm BA)} (\vec k) = \begin{pmatrix}
\frac{\Delta}{2} & -\gamma_0 f(\vec k) & 0 &\gamma_1 \\ 
-\gamma_0 f^* (\vec k) & \frac{\Delta}{2} &   -\gamma_3 f (\vec k)  &  0 \\
0 &-\gamma_3 f^* (\vec k)  & -\frac{\Delta}{2} & -\gamma_0 f(\vec k) \\
 \gamma_1& 0 & -\gamma_0 f^* (\vec k) & -\frac{\Delta}{2}
\end{pmatrix}. 
\ee

We note that for the Hamiltonians above, the band extrema occur at the corners of the Brillouin zone labeled by $\vec K_+ = (\frac{4 \pi}{3a}, 0)$ and $\vec K_- = -\vec K_+$. 
At low energy, the tight-binding Hamiltonian can be approximated up to the linear order of $\vec p = \vec k - \vec K_\xi$ around each valley $\xi = \pm$. By expanding $f (\vec k)$ around $\vec K_\xi$, we obtain
\be
H^{\rm (AB)} (\vec p) = \begin{pmatrix}
\frac{\Delta}{2} & \hbar v p^\dag & 0 & \hbar v_3 p \\ 
\hbar v p & \frac{\Delta}{2} & \gamma_1  &  0 \\
0 & \gamma_1 & -\frac{\Delta}{2} & \hbar v p^\dag \\
\hbar v_3 p^\dag & 0 & \hbar v p & -\frac{\Delta}{2}
\end{pmatrix}, \quad 
H^{\rm (BA)} (\vec p) = \begin{pmatrix}
\frac{\Delta}{2} & \hbar v p^\dag & 0 &\gamma_1 \\ 
\hbar v p & \frac{\Delta}{2} &  \hbar v_3 p^\dag   &  0 \\
0 &  \hbar v_3 p & -\frac{\Delta}{2} &\hbar v p^\dag \\
\gamma_1 & 0 & \hbar v p & -\frac{\Delta}{2}
\end{pmatrix},
\ee
where $p = \xi p_x + i p_y$, $v = \sqrt{3} a \gamma_0 / 2\hbar$  is the Fermi velocity in each layer and $v_3 =  \sqrt{3} a \gamma_3 / 2\hbar$.

\section{Numerical Calculation of Shift Vector in AB stacked BLG for other polarizations} 
\begin{figure}[h]
    \centering
    \includegraphics[scale=0.4]{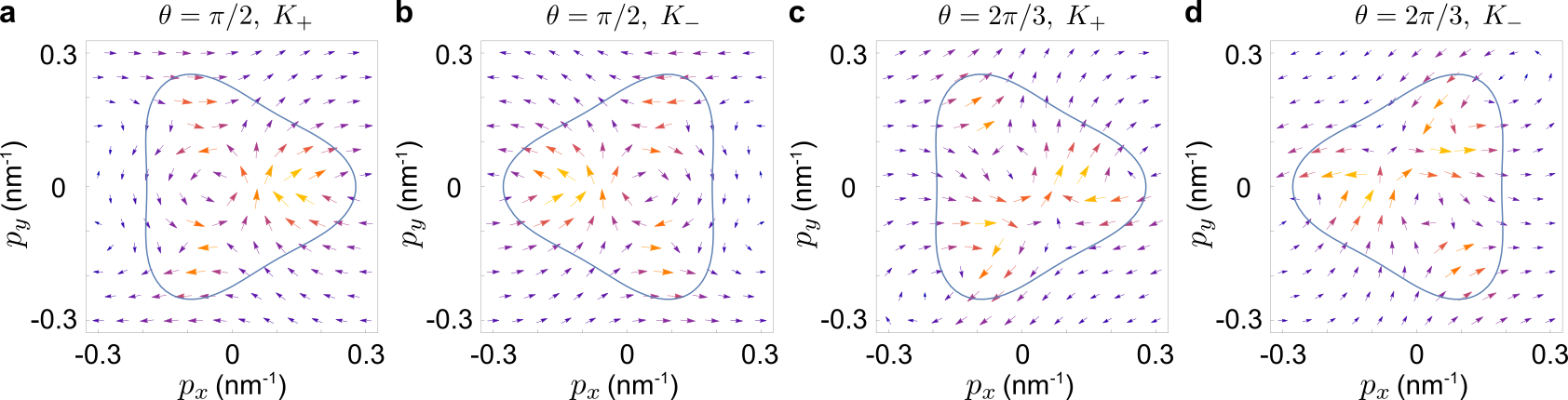}
    \caption{(a,b) Plot of $\vec R^{(\rm AB)} (\Delta,\theta= \pi/2, \vec p)$ at the $K_+$ (a) and $K_-$ (b) valley. (c,d) Plot of $\vec R^{(\rm AB)} (\Delta, \theta = 2\pi/3, \vec p)$ at the $K_+$ (c) and $K_-$ (d) valley. Similar to that found in the main text, we have taken $\Delta = 20 \, {\rm meV}$. All other parameters are the same as Fig.~2 in the main text.}
    \label{figS3}
\end{figure}

Using the four-band Hamiltonian derived in the previous section, we numerically calculate the weighted shift vectors $\vec R^{(\eta)} (\Delta, \theta, \vec p)$ in $k$-space in the vicinity of the Dirac point. In the main text, we have shown $\vec R^{(\eta)} (\Delta, \theta, \vec p)$ in AB/BA stacked BLG for the $x$-polarised electric at the $K_+$ valley. Here we plot the $\vec R^{(\rm AB)} (\Delta, \theta, \vec p)$ at $\theta = \pi/2$ (Fig.~\ref{figS3}a and b) and $\theta = 2 \pi/3$ (Fig.~\ref{figS3}c and d) at $K_+$ and $K_-$ valleys. 

We note that for $\vec K_\pm = \left( \pm \frac{4 \pi}{3 a}, 0 \right)$ and $\vec k $ measured from the $\Gamma$ point, $k_x \to - k_x$ maps $p_x$ at the $K_\pm$ valley to $- p_x$  at the $K_\mp$ valley, while $k_y \to - k_y$ maps $p_y \to -p_y$ at the same valley. Using these identities, we observe that for the electric field polarisation along with the mirror plane ($y$-axis, $\theta = \pi/2$), the shift vectors in Fig.~\ref{figS3}a and b obey the relations in Eq.~(\ref{Seq:My}) and~(\ref{Seq:TMy}). From $p_x$ at the $K_+$ valley to $-p_x$ at the $K_-$ valley, $R_x^{(\rm AB)}$ switches sign and $R_y^{(\rm AB)}$ remains unchanged. For $p_y \to - p_y$ in each valley, $R_x^{(\rm AB)}$ is also odd while $R_y^{(\rm AB)}$ is even. This leads to a shift photocurrent flowing along the $y$ axis. 

Furthermore, when the electric field is perpendicular to one of the reflection axis ($\theta = 2 \pi/3$), we observe that the weighted shift vector component normal to the polarisation is even, while the component parallel to the polarisation is odd (e.g. Fig.~\ref{figS3}c and d), leading to a transverse shift photocurrent.

\end{document}